\documentclass[conference]{IEEEtran}
\IEEEoverridecommandlockouts
\usepackage{cite}
\usepackage[normalem]{ulem}
\usepackage{amsmath,amssymb,amsfonts}
\usepackage{algorithmic}
\usepackage{graphicx}
\usepackage{textcomp}
\usepackage{xcolor}
\usepackage{caption}
\usepackage{subcaption}  %subfigure
\usepackage{url}
\usepackage{pifont}
\usepackage{colortbl}
\usepackage{hyperref}
\usepackage{enumitem}
\usepackage{extarrows} 
\usepackage{tikz}
\usepackage{tabularx}
\usepackage{stmaryrd}
\usepackage{makecell}
\usepackage[ruled,vlined,noend,linesnumbered]{algorithm2e}
\usepackage{adjustbox}
\usepackage{soul}
\usepackage{balance}
\usepackage{booktabs}
\usepackage{colortbl}
\usepackage{multirow}
\usepackage{arydshln}
\usepackage[misc]{ifsym}
\usepackage[draft]{fixme} %draft -> final
\fxsetup{layout=footnote, theme=color} %marginclue
\definecolor{fxnote}{rgb}{0.8000,0.0000,0.0000}

\renewcommand{\thetable}{\arabic{table}}
\usepackage{pgfplots}
\usepgfplotslibrary{groupplots}
\usepgfplotslibrary{statistics}
\usetikzlibrary{patterns}
\pgfplotsset{width=8cm, compat=1.9, max space between ticks=35}
\definecolor{darkred}{rgb}{0.70, 0, 0}
\definecolor{darkgreen}{rgb}{0, 0.55, 0}
\definecolor{darkblue}{rgb}{0, 0.0, 0.78}
\definecolor{darkpurple}{rgb}{0.53, 0, 0.50}
\definecolor{purple}{rgb}{0.57, 0.55, 0.78}
\definecolor{iryellow}{rgb}{0.66, 0.82, 0.56}
\definecolor{trolleygrey}{rgb}{0.5, 0.5, 0.49}
\definecolor{tropicalrainforest}{rgb}{1.0, 0.91, 0.7}
\definecolor{glaucous}{rgb}{0.38, 0.51, 0.71}
\definecolor{cardinal}{rgb}{0.7, 0.33, 0.3}
\definecolor{palegreen}{rgb}{0.61, 0.7, 0.35}
\definecolor{pink}{rgb}{0.97, 0.78, 0.65}
\definecolor{orange}{rgb}{0.96, 0.69, 0.51}
\definecolor{purple}{rgb}{0.57, 0.55, 0.78}

\def\BibTeX{{\rm B\kern-.05em{\sc i\kern-.025em b}\kern-.08em
    T\kern-.1667em\lower.7ex\hbox{E}\kern-.125emX}}

\begin{document}

\title{PURPLE: Making a Large Language Model a Better SQL Writer}

\author{
\IEEEauthorblockN{
Tonghui Ren{$^\dagger$}, 
Yuankai Fan{$^\dagger$}, 
Zhenying He{$^\dagger$}, 
Ren Huang{$^\dagger$}, 
Jiaqi Dai{$^\dagger$}, 
Can Huang{$^\dagger$},  \\
Yinan Jing{$^\dagger$}, 
Kai Zhang{$^\dagger$}, 
Yifan Yang{$^\ddagger$},
X.Sean Wang{$^\dagger$}
}
\IEEEauthorblockA{
\textit{{$^\dagger$}School of Computer Science, Fudan University} \\ 
\textit{{$^\ddagger$}Transwarp Technology (Shanghai) Co., Ltd}
}
% Shanghai, China \\
thren22@m.fudan.edu.cn, {\{fanyuankai, zhenying\}}@fudan.edu.cn,  {\{renhuang21, daijq22, huangcan22\}}@m.fudan.edu.cn, \\
{\{jingyn, zhangk\}}@fudan.edu.cn, yifan.yang@transwarp.io, xywangCS@fudan.edu.cn

}
\maketitle

\begin{abstract}

Large Language Model (LLM) techniques play an increasingly important role in Natural Language to SQL (NL2SQL) translation.
LLMs trained by extensive corpora have strong natural language understanding and basic SQL generation abilities without additional tuning specific to NL2SQL tasks.
Existing LLMs-based NL2SQL approaches try to improve the translation by enhancing the LLMs with an emphasis on user intention understanding.
However, LLMs sometimes fail to generate appropriate SQL due to their lack of knowledge in organizing complex logical operator composition.
A promising method is to input the LLMs with demonstrations, which include known NL2SQL translations from various databases. LLMs can learn to organize operator compositions from the input demonstrations for the given task.
In this paper, we propose PURPLE (Pre-trained models Utilized to Retrieve Prompts for Logical Enhancement), which improves accuracy by retrieving demonstrations containing the requisite logical operator composition for the NL2SQL task on hand, thereby guiding LLMs to produce better SQL translation.
PURPLE achieves a new state-of-the-art performance of 80.5\% exact-set match accuracy and 87.8\% execution match accuracy on the validation set of the popular NL2SQL benchmark Spider.
PURPLE maintains high accuracy across diverse benchmarks, budgetary constraints, and various LLMs, showing robustness and cost-effectiveness.

\end{abstract}


% ============================ Introduction ============================
\section{Introduction}

The task of Natural Language to SQL (NL2SQL) translation helps the Database Management Systems (DBMS) be more user-friendly. 
The NL2SQL approach translates Natural Language (NL) query into SQL based on the database, enabling users to easily access data in a DBMS without needing knowledge of the database schema or SQL syntax.

Recently, general-purpose Large Language Models (LLMs) have exhibited profound capabilities in various downstream tasks without the need for a costly LLM fine-tuning process, including NL2SQL~\cite{DBLP:Synchromesh, DBLP:DIN-SQL, DBLP:SQL-PALM, DBLP:EvaluatingLLM, DBLP:ChatGPT-zero-shot, DBLP:SELF-DEBUG, DBLP:ZERO, DBLP:DAIL-SQL}.
Thanks to the strong NL understanding ability, existing approaches can achieve high Execution Match\footnote{Execution Match: SQL equivalence based on the execution result.} accuracy.
For example, DIN-SQL~\cite{DBLP:DIN-SQL} is one of the state-of-the-art (SOTA) approaches based on a few-shot Chain-of-Thought (CoT) strategy~\cite{DBLP:COT},
which can achieve 82.8\% execution match accuracy on the validation set of the NL2SQL benchmark, Spider~\cite{DBLP:Spider}.

\begin{table}[htbp]
    \setlength\tabcolsep{6pt}
    \centering
    \caption{LLMs-based approaches accuracy on Spider.}
    % \vspace{-3mm}
    \begin{tabular}{l c c}
        \hline
        \textbf{Strategy}        & \textbf{Exact-Set Match\%} & \textbf{Execution Match\%} \\ 
        \hline
        ChatGPT-SQL           &     37.9 &     70.1  \\
        C3                    &     43.1 &     81.8  \\
        DIN-SQL(GPT4)         &     60.1 &     82.8  \\
        DAIL-SQL(GPT4)        &     \textbf{68.7} &     83.6  \\
        \hline
        \end{tabular}
    \label{tab:intro_cmp}
    \vspace{-3mm}
\end{table}

Upon analyzing the translations of existing LLMs-based approaches, we observe that they achieve high execution accuracy thanks to the strong NL understanding ability of the LLMs, while the LLMs only have basic SQL knowledge for SQL writing. 
We notice that all of the existing LLMs-based NL2SQL approaches fail to achieve high Exact-Set Match\footnote{Exact-Set Match: SQL equivalence at the SQL component level.} accuracy as shown in Table~\ref{tab:intro_cmp}, which is more rigorous compared to execution match accuracy. The SQL queries with the same execution result may have different semantics, which means execution match accuracy will overestimate the performance of approaches, leading to the false positive~\cite{DBLP:Spider}.
The complexity of SQL is mainly from the \textbf{logical operator composition}, which is not what general LLMs are good at.

\begin{figure}
 \begin{subfigure}[b]{1.0\linewidth}
 \centering
 \includegraphics[width=\linewidth]{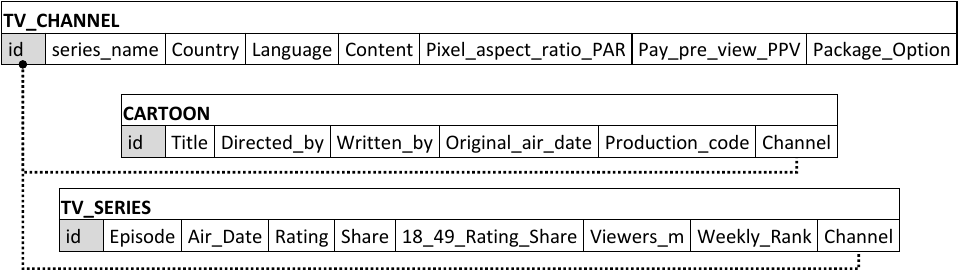}
 % \vspace{-43mm}
 \caption{Database schema from Spider for the example.}
 \vspace{5mm}
 \label{fig:intro_example_db}
 \end{subfigure}
 
 \begin{subfigure}[b]{1.0\linewidth}
  \centering
   \begin{adjustbox}{width=\textwidth}
   % \begin{tabular}{l l} \Xhline{3\arrayrulewidth}
   \begin{tabular}{l l} \toprule
     NL \;\;Query: & \emph{\makecell[l]{\fontsize{11pt}{13pt}\selectfont What are the \textbf{countries} that are not playing \textbf{cartoons} written by \textbf{Todd Casey}?}}  \\
     \hline \hline & \\[-2.0ex]

     & \makecell[c]{C3 (ChatGPT)} \\
     \makecell[l]{Generated SQL: \\ \textcolor{darkred}{\textbf{(Incorrect)}} \\ \;} & \makecell[l]{
     {\fontfamily{pcr}\selectfont SELECT TV\_CHANNEL.Country FROM TV\_CHANNEL WHERE }
     \\ {\fontfamily{pcr}\selectfont \textcolor{darkred}{\textbf{TV\_CHANNEL.id NOT IN (SELECT CARTOON.channel FROM CARTOON}}}
     \\ {\fontfamily{pcr}\selectfont WHERE CARTOON.Written\_by=`Todd Casey'\textcolor{darkred}{\textbf{)}};}
     } \\[-2.0ex] \\ 
     
     \hline & \\[-2.0ex]
     
     & \makecell[c]{DAIL-SQL (GPT4)} \\
     \makecell[l]{Generated SQL: \\ \textcolor{darkred}{\textbf{(Incorrect)}} \;} & \makecell[l]{
     {\fontfamily{pcr}\selectfont SELECT Country FROM TV\_CHANNEL WHERE \textcolor{darkred}{\textbf{id NOT IN (}}}
     \\ {\fontfamily{pcr}\selectfont \textcolor{darkred}{\textbf{SELECT Channel FROM CARTOON}} WHERE Written\_by=`Todd Casey'\textcolor{darkred}{\textbf{)}};}
     }\\[-2.0ex] \\ 
     
     \hline & \\[-2.0ex]

     & \makecell[c]{DIN-SQL (GPT4)} \\
     \makecell[l]{Generated SQL: \\ \textcolor{darkred}{\textbf{(Incorrect)}} \\ \;} & \makecell[l]{
     {\fontfamily{pcr}\selectfont SELECT  Country FROM TV\_CHANNEL WHERE Country \textcolor{darkred}{\textbf{NOT IN (}}}
     \\ {\fontfamily{pcr}\selectfont SELECT T1.Country FROM  TV\_CHANNEL AS T1 JOIN CARTOON AS T2}
     \\ {\fontfamily{pcr}\selectfont ON T1.id = T2.Channel WHERE  T2.Written\_by=`Todd Casey'\textcolor{darkred}{\textbf{)}};}
     }\\[-2.0ex] \\ 
     
     \hline & \\[-2.0ex]
     
     \makecell[l]{Gold SQL: \\ \\ \\ \;} & \makecell[l]{
     {\fontfamily{pcr}\selectfont SELECT Country FROM TV\_CHANNEL }
     \\ {\fontfamily{pcr}\selectfont \textcolor{darkgreen}{\textbf{EXCEPT}}}
     \\ {\fontfamily{pcr}\selectfont \textcolor{darkgreen}{\textbf{SELECT T1.Country FROM TV\_CHANNEL AS T1 JOIN CARTOON AS T2}}}
     \\ {\fontfamily{pcr}\selectfont \textcolor{darkgreen}{\textbf{ON T1.id = T2.Channel}} WHERE T2.Written\_by=`Todd Casey';}
     }
     \\[-2.0ex]  \\ \Xhline{3\arrayrulewidth}
   \end{tabular}
  \end{adjustbox}
  \caption{NL query from Spider and the corresponding translation result from different approaches.}
  
  \label{fig:intro_example_task}
 \end{subfigure}
 \caption{An example of NL2SQL translation task from Spider.}
 \vspace{-3mm}
 \label{fig:intro_example}
\end{figure}

Existing works apply zero-shot or few-shot strategies to enhance LLMs with task-specific knowledge. Zero-shot approaches, such as C3~\cite{DBLP:C3}, employ instructional prompts to guide the utilization of SQL keywords. 
On the other hand, DAIL-SQL~\cite{DBLP:empowerd} and DIN-SQL~\cite{DBLP:DIN-SQL} are two few-shot strategies to improve the capabilities of LLMs through few-shot learning~\cite{DBLP:llm_few_shot_learner}.
Both DAIL-SQL and DIN-SQL emphasize the importance of NL understanding. 
DAIL-SQL integrates SQL keyword similarity for demonstration selection.
However, they fail to provide the knowledge of operator composition in SQL formulation.
The LLMs understand the user intention but lack related knowledge in organizing logical operator composition for SQL generation, resulting in a semantic similar but incorrect SQL.
Figure~\ref{fig:intro_example} illustrates a case of NL2SQL task. The abovementioned approaches implement the ``\textit{NOT IN}'' operator, corresponding to the ``\textit{not playing}'' in the NL query. Both C3 and DAIL-SQL do capture the meaning of excluding ``\textit{countries}'' but failed to implement it because such a semantic needs a ``\textit{JOIN}'' operator in the SQL. DIN-SQL generates a likely correct prediction but fails to recognize that the ``\textit{EXCEPT}'' keyword involves a de-duplication operation, resulting in redundant outcomes.
\textbf{Despite the three SOTA LLM-based approaches capturing user intentions, they failed in managing complex logical operator compositions.} Such as the necessity for a ``\textit{JOIN}'' operator or in distinguishing the difference between ``\textit{NOT IN}'' and ``\textit{EXCEPT}'' in SQL.

In this study, we aim to enhance the SQL generation capabilities of general LLMs on NL2SQL tasks, making an LLM a better SQL writer.
We hope that such an approach can achieve high execution accuracy by leveraging the robust NL comprehension inherent to LLMs, as well as high exact-set match accuracy to maintain logical semantic integrity. 
The \textbf{main challenge} is to provide requisite logical composition knowledge without exceeding the input length budget. Given the limited input length and the infinite potential logical compositions, it is impractical to contain all composition knowledge within the prompt.

To enhance the LLMs with corresponding SQL logical operator composition knowledge within the limited input length, we introduce PURPLE, Pre-trained models Utilized to Retrieve Prompts for Logical Enhancement, a novel few-shot prompting strategy tailored for LLMs-based NL2SQL translation.
The key point of PURPLE is the \textbf{demonstration selection}, which needs to select the demonstrations containing the requisite logical operator composition.
The demonstrations\footnote{A detailed description of demonstrations is shown in Section~\ref{subsec:demonstrations}.} are NL2SQL tasks derived from various databases, each containing an NL query, database information, and SQL translation.
LLMs can learn from the demonstrations in the prompt, which involves the selected demonstrations and the description of the current NL2SQL task, about how to handle the NL2SQL task, especially managing the operator composition, which is challenging for LLMs.
We employ a fine-tuned model to identify the logical operator compositions knowledge relevant to the current task.
Moreover, we introduce a demonstration selection strategy based on the inferred knowledge. This approach is for both generalization and fuzzification, considering the limited size of all demonstrations and the capabilities of the fine-tuned prediction model.

PURPLE consists of four main modules: \textbf{Schema Pruning}, \textbf{Skeleton Prediction}, \textbf{Demonstration Selection}, and \textbf{Database Adaption}. Initially, PURPLE employs a classifier and a probability-based algorithm to prune irrelevant schema items for a given NL query.
Subsequently, the pruned schema is used to infer a SQL skeleton, which masks all database-specific values compared with SQL, that contains the requisite operator composition knowledge.
PURPLE retrieves relevant demonstrations based on the inferred skeleton.
Following the LLM calling, PURPLE adjusts the output to adapt to the specific database schema and SQL dialect, thereby mitigating the LLM-induced hallucination problems.

To show the performance of PURPLE, we conduct a comprehensive evaluation of our strategy from multiple perspectives on four mainstream benchmarks. Moreover, we explore the trade-off between cost and performance. Notably, PURPLE is flexible because it can be configured for higher performance at a higher cost or optimized for reducing the expense of some performance drop. We further compare various approaches across different LLMs to evaluate the performance fluctuation. An ablation study is also conducted to show the effectiveness of each module in PURPLE.

The contributions of this paper are summarized as follows:

\begin{itemize}
 \item We propose PURPLE, a novel approach leveraging pre-trained models to generate optimized prompts for LLMs and augment the performance of NL2SQL translation.
 \item We enhance the SQL writing ability of LLMs by selecting demonstrations containing the requisite operator composition knowledge, helping the LLMs perform better.
 \item We conceptualize the SQL logical composition knowledge through an automaton framework, defining four levels of automaton state abstraction. This modeling helps select the valuable demonstration for PURPLE. 
 \item We test PURPLE through comprehensive experiments. The outcomes show superior performance, especially an 11.8\% improvement in exact-set match accuracy compared to the existing LLMs-based NL2SQL approaches. The experiment also shows the robustness and cost-effectiveness of PURPLE.
\end{itemize}

The paper is organized as follows: Section~\ref{sec:preliminaries} introduces essential preliminaries. Section~\ref{sec:method_overview} gives an overview of PURPLE. The core modules are explained in Section~\ref{sec:methodologies}. Experimental results are discussed in Section~\ref{sec:experiments}. Related works and conclusions are shown in Section~\ref{sec:related_works} and Section~\ref{sec:conclusion_and_future_work}.

% ============================ PRELIMINARIES ============================
\section{PRELIMINARIES}\label{sec:preliminaries}

NL2SQL translation has benefited from advancements in Natural Language Processing (NLP). This section provides the foundational concepts and definitions relevant to this study.

\subsection{Language Models}

A language model (LM) is a statistical model fundamental to many NLP tasks, typically trained on extensive text corpora.
LMs have been applied to various tasks, including NL2SQL. We categorize LMs as PLMs and LLMs in this paper.

\textbf{PLMs:} PLMs refer to LMs with a relatively smaller parameter size in this paper, which can not applied to downstream tasks without fine-tuning. Limited by their model size and pre-training corpus, these models do not exhibit capabilities that can be directly applied to downstream tasks. Notable examples of PLMs include BERT~\cite{DBLP:BERT}, BART~\cite{DBLP:BART}, and T5~\cite{DBLP:T5}.

\textbf{LLMs:} This category refers to LMs with a huge parameter size demonstrating ability across many downstream tasks without tuning. Instruction design or in-context learning can be employed to adapt LLMs to different tasks. Such models include GPT3~\cite{DBLP:GPT3}, PaLM~\cite{DBLP:PALM}, ChatGPT, and GPT4.

\subsection{LLMs-based NL2SQL}

The abilities of LLMs for NL understanding and generation have drawn attention from researchers.
We categorize LLMs-based NL2SQL approaches into zero-shot and few-shot.
Both approaches enhance the performance of LLMs on downstream tasks through prompts, which are sequences of textual instructions that elicit outputs from LLMs.

For a typical NL2SQL translation task, the input consists of an NL query $\mathcal{X}$ and database information $\mathcal{D}$. The goal is to obtain the target SQL $\mathcal{Y}$. The task can be formulated as:
$$\mathcal{\hat{Y}}=LLM(P(\mathcal{X}, \mathcal{D}, \mathcal{E}))$$
In this function, $LLM$ denotes the LLM call, $P$ denotes the prompt generation, and $\mathcal{E}$ denotes known NL2SQL translation that can be used as auxiliary information for the LLMs. 

\textbf{Zero-shot:} Zero-shot NL2SQL translation does not include annotated examples. In this context, $P$ can be represented as:
$$P_0(\mathcal{X}, \mathcal{D}, \mathcal{\varnothing})$$ 
The prompt generation relies solely on the information of the current translation task. C3~\cite{DBLP:C3} and ChatGPT-SQL~\cite{DBLP:ChatGPT-zero-shot} are two typical zero-shot NL2SQL approaches.

\textbf{Few-shot:} With a few demonstrations from the annotated datasets, the LLMs can learn how to generate the correct SQL. We consider the training set of the benchmark as the source of demonstrations, maintaining the cross-domain setting. The prompt generation process can be represented as:
$$P_f(\mathcal{X}, \mathcal{D}, \mathcal{E})$$
The $\mathcal{E}$ is the demonstrations from annotated datasets, detailed descriptions will be formally outlined in section~\ref{subsec:demonstrations}.

\subsection{SQL skeleton}\label{subsec:sql_skeleton}

In this study, we introduce the concept of a SQL skeleton, denoted as $\mathcal{S}$. The skeleton serves as a structural template abstracting from database-specific details, thereby focusing on the logical operator composition inherent within SQL queries. The skeleton preserves all operational keywords while substituting placeholders for specific database elements like tables, columns, and constant values. For instance, the SQL skeleton of the gold SQL in Figure~\ref{fig:intro_example_task} is:

{\footnotesize \fontfamily{pcr}\selectfont \textcolor{darkpurple}{\textbf{SELECT}}\;\_\;\textcolor{darkpurple}{\textbf{FROM}}\;\_}

{\footnotesize \fontfamily{pcr}\selectfont \textcolor{darkpurple}{\textbf{EXCEPT}}}

{\footnotesize \fontfamily{pcr}\selectfont \textcolor{darkpurple}{\textbf{SELECT}}\;\_\;\textcolor{darkpurple}{\textbf{FROM}}\;\_\;\textcolor{darkpurple}{\textbf{JOIN}}\;\_\;\textcolor{darkpurple}{\textbf{ON}}\;\_\;\textcolor{darkpurple}{\textbf{=}}\;\_\;\textcolor{darkpurple}{\textbf{WHERE}}\;\_\;\textcolor{darkpurple}{\textbf{=}}\;\_}

\noindent This abstraction focuses on the operational logic of the SQL, providing a generalized yet structurally representative form.

% ============================ Method Overview ============================
\section{Method Overview}\label{sec:method_overview}

In this section, we explore the demonstrations as an input source for PURPLE and present an overview of the pipeline.

\subsection{Demonstration}\label{subsec:demonstrations}

\begin{figure}
 \begin{subfigure}[b]{1.0\linewidth}
 \centering
 \includegraphics[width=\linewidth]{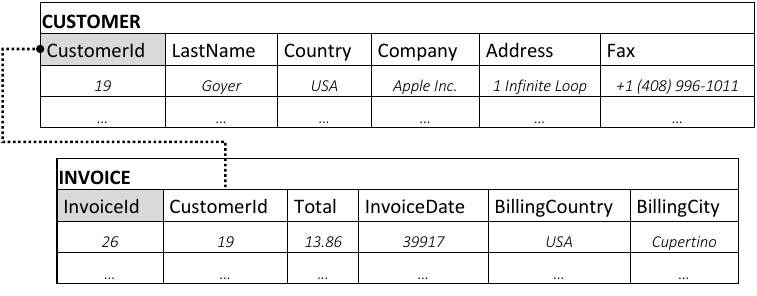}
 \caption{Database information for a demonstration}
 \vspace{4mm}
 \label{fig:demonstration_schema}
 \end{subfigure}
 \begin{subfigure}[b]{1.0\linewidth}
  \centering
  \begin{adjustbox}{width=\textwidth}
   \begin{tabular}{c l} \Xhline{3\arrayrulewidth}
     NL: & \emph{\makecell[l]{What are the last names of customers without invoice totals exceeding 20?}}  \\
     \hline
     \makecell[c]{SQL: \\ \textcolor{darkpurple}{\textbf{(Skeleton)}}} & \makecell[l]{
     {\fontfamily{pcr}\selectfont \textcolor{darkpurple}{\textbf{SELECT}} LastName \textcolor{darkpurple}{\textbf{FROM}} CUSTOMER}
     \\ {\fontfamily{pcr}\selectfont \textcolor{darkpurple}{\textbf{EXCEPT}}}
     \\ {\fontfamily{pcr}\selectfont \textcolor{darkpurple}{\textbf{SELECT}} T1.LastName \textcolor{darkpurple}{\textbf{FROM}} CUSTOMER AS T1 \textcolor{darkpurple}{\textbf{JOIN}} Invoice AS}
     \\ {\fontfamily{pcr}\selectfont T2 \textcolor{darkpurple}{\textbf{ON}} T1.CustomerId \textcolor{darkpurple}{\textbf{=}} T2.CustomerId \textcolor{darkpurple}{\textbf{WHERE}} T2.total \textcolor{darkpurple}{\textbf{>}} 20}
     } \\ \Xhline{3\arrayrulewidth}
   \end{tabular}
  \end{adjustbox}
  \caption{NL query and SQL for a demonstration}
  \label{fig:demonstration_nl_sql}
 \end{subfigure}
 \caption{An example for demonstrations}
 \label{fig:demonstration}
 \vspace{-3mm}
\end{figure}

In the context of few-shot LLMs-based NL2SQL translation, demonstrations serve as examples that LLMs can learn to handle the current task. Each demonstration consists of task inputs and corresponding outputs. Based on cross-database settings in this paper, we employ the original training data from the NL2SQL benchmarks as demonstrations.

Specifically, a demonstration $e_i \in \mathcal{E}$ contains three components: the NL $\mathcal{X}^{e_i}$, the database $\mathcal{D}^{e_i}$, and the target SQL $\mathcal{Y}^{e_i}$. 
Figure~\ref{fig:demonstration} provides an illustrative example of a demonstration. The $\mathcal{D}^{e_i}$ includes the database schema and the data. We select a subset of representative values for each column like~\cite{DBLP:BRIDGE} to optimize the length of database information. Formally, a demonstration can be represented as:
$$e_i = \mathrm{CAT}(\mathcal{D}^{e_i}, \mathcal{X}^{e_i}, \mathcal{Y}^{e_i})$$
Here, $\mathrm{CAT}$ represents the string concatenation. The prompt structure within PURPLE is formulated as:
$$P_f = \mathrm{CAT}(\mathcal{E}', \mathcal{D}, \mathcal{X})$$
In this expression, $\mathcal{E}'$ represents the subset of demonstrations selected from the entire set $\mathcal{E}$ for constructing the prompt.

Moreover, we incorporate a schema pruning strategy in PURPLE. Accordingly, the schema of each demonstration undergoes a pruning process to reduce its length. Section~\ref{sec:schema_pruning} provides details of such a module. Thus, the database information for a demonstration is a subset of the whole database.

\subsection{Overview of PURPLE}

\begin{figure}
  \centering
  \includegraphics[width=0.8\linewidth]{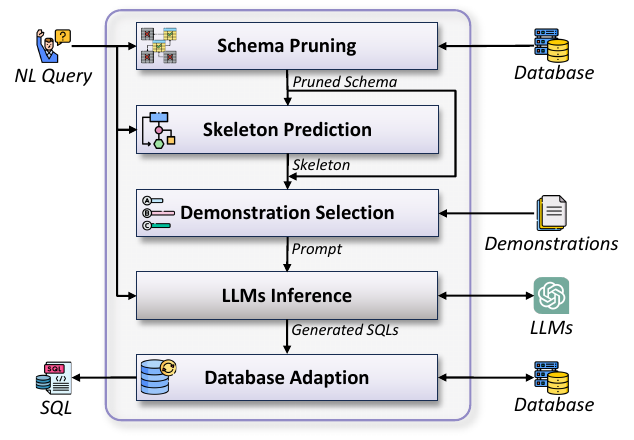}
  \caption{Overview of PURPLE.}
  \label{fig:purple_overview}
  \vspace{-3mm}
\end{figure}

The architecture of PURPLE is shown in Figure \ref{fig:purple_overview}. 
Firstly, the schema pruning module of PURPLE excludes tables and columns that are not requisite for constructing the target SQL for the current NL query.

The pruned schema and the NL query are used for SQL skeleton prediction. 
Such a SQL skeleton represents the needed logical composition knowledge required by LLMs.
PURPLE selects relevant examples based on the skeleton to form a prompt, which also includes the NL query and the pruned schema for the current NL2SQL task.

PURPLE submits the prompt to LLMs for NL2SQL translation. A database adaption module follows, detecting and fixing hallucination errors induced by the LLMs. PURPLE integrates an execution-consistency~\cite{DBLP:consistency} strategy into the database adaption module to stabilize the output further. The resulting processed SQL becomes the final output of PURPLE.

\subsubsection{Schema Pruning}

As Step 1 of Figure \ref{fig:purple_overview} illustrates, schema pruning narrows down the database information. This module decides which tables or columns are needed for the target SQL based on the NL query and schema. This step prunes the schema to reduce inference complexity for higher translation accuracy, as subsequent modules only process the pruned schema. 
It is important to keep high recall to reduce the risk of error propagation.
We design a pruning strategy based on a trained classifier, trying to keep essential tables or columns while keeping the database information short.

\subsubsection{Skeleton Prediction}

Step 2 of Figure \ref{fig:purple_overview} is the skeleton prediction module, which detects the requisite logical composition knowledge for the NL2SQL task. Accurate predictions allow us to extract demonstrations containing essential knowledge. 
We employ a specialized fine-tuned PLM on the skeleton generation task.
The fine-tuning phase equips the PLM with the capability to discern operator compositions.
We generate the top-$k$ skeletons by the beam search for high recall.

\subsubsection{Demonstration Selection}

Highlighted in Step 3 of Figure \ref{fig:purple_overview}, 
PURPLE selects demonstrations following the predicted SQL skeletons. While it is non-trivial to model the composition knowledge and extract the demonstrations based on the predicted requisite.
The selection strategy must have the capacity for generalization to address unseen tasks and incorporate fuzzification to compensate for the limitations of the skeleton prediction model. The complexity of composition knowledge cannot be captured by simplistic similarity functions.
We design four levels of SQL skeleton abstraction to facilitate the selection of demonstrations that include composition knowledge for the LLMs.
Each higher abstraction level masks more details, focusing on more coarse-grained composition.
Such an approach significantly enhances the generalization capabilities of PURPLE for unseen logical composition.

\subsubsection{Database Adaption}

Step 5 in Figure \ref{fig:purple_overview} presents the database adaption module.
Hallucination issues in LLMs are a common occurrence, often resulting in the generation of buggy SQL queries that are incompatible with specific databases. Unlike methods such as PICARD~\cite{DBLP:PICARD} that employ specialized decoding strategies, we face challenges since we use LLMs as a service.
To reduce the buggy SQL generation, we systematically catalog these errors and develop heuristic-based correction algorithms to address them, a low-cost strategy helping LLMs correct the buggy SQL. Such a process can make the output of LLMs fit specific databases, including the specific database schema and specific DBMS SQL dialect. We also include an execution-consistency strategy into PURPLE to stabilize the LLMs generation.

\section{METHODOLOGIES}\label{sec:methodologies}

\subsection{Schema Pruning}\label{sec:schema_pruning}

PURPLE begins with a Schema Pruning module, which can be used to eliminate the schema items that will not be used in the target SQL.
This module introduces two benefits: Firstly, it shortens the input length for each demonstration, enabling more demonstrations within the token input constraint. 
Secondly, it simplifies the inference task for LLMs by limiting the problem to a subset of the database schema.

\subsubsection{Table-Column Classifier}\label{subsec:table_column_classifier}

The module takes as input the schema denoted by $\mathcal{D}=<\mathcal{T}, \mathcal{C}, \mathcal{P}, \mathcal{F}>$ and NL query denoted by $\mathcal{X}$. More specifically, $\mathcal{T}=\{t_1, t_2, ..., t_{|\mathcal{T}|}\}$ denote the tables within the database schema. For each table $t_i$, the columns are denoted by $\mathcal{C}_i=\{c_{i,1}, c_{i, 2}, ..., c_{i, {|\mathcal{C}_i|}}\}$. $\mathcal{P}=\{c_{p_1}, c_{p_2}, ..., c_{p_{|\mathcal{P}|}}\}$ represents the primary keys, and $\mathcal{F}=\{(c_{f_1}, c_{p_1}),(c_{f_2}, c_{p_2}),...,(c_{f_{|\mathcal{F}|}}, c_{p_{|\mathcal{F}|}})\}$ represents the foreign-primary key pairs.

We implement such a classifier based on the schema ranking module of RESDSQL~\cite{DBLP:RESD}. The input can be structured as:
$$\mathcal{X} , t_1 , c_{1, 1}, ..., c_{1,{|\mathcal{C}_1|}} , ... , t_{|\mathcal{T}|} , c_{{|\mathcal{T}|}, 1}, ..., c_{{|\mathcal{T}|}, {|\mathcal{C}_{|\mathcal{T}|}|}}$$
For each $t_i$ and $c_{i, j}$, the classifier will predict whether such table or column is related to the question.

The classifier is trained by the NL2SQL training data. For each input pair of $(\mathcal{X}, \mathcal{D})$, the labels are extracted from the SQL $\mathcal{Y}$ to identify the presence (absence) of each table or column. Training adopts focal loss~\cite{DBLP:focal_loss} in line with RESDSQL.

In the inference stage, the classifier yields the probability of relevance for each schema item to the NL query. 
Tables with a probability exceeding the threshold $\tau_p$ are denoted as $\mathcal{T}'$. Similarly, for each table $t_i$, columns with a probability exceeding $\tau_p$ are denoted as $\mathcal{C}_{i}'$.

PURPLE adopts a novel method of schema pruning, distinct from the existing methods.
The conventional strategy retains the top-$k_1$ tables and top-$k_2$ columns, leading to two disadvantages. Firstly, it tends to increase the complexity of schema by including superfluous schema items. Secondly, the selected tables may lack connectivity due to the limited precision of the classifier. These factors necessitate additional processing by LLMs to differentiate among an expanded set of schema items.
In contrast, we aim to identify a schema subset that is both closely related and interconnected. 
PURPLE models the schema pruning task within the framework of a \textit{Steiner Tree Problem}~\cite{DBLP:steiner_tree},
similar to the keywords search studies~\cite{DBLP:DISCOVER}.
We include a redundant boundary to optimize recall.

We represent the schema as a graph $G=(V, E)$, with $V$ representing tables $\mathcal{T}$, and $E$ representing the relationships between them (foreign-primary key connections). Each edge in $E$ is assigned a weight of 1. The tables in $\mathcal{T}'$ as shown in Section~\ref{subsec:table_column_classifier} can be reduced to the Steiner point set $S$. So the pruning strategy can be reduced to the \textit{Steiner Tree Problem}, the objective is to extract the smallest connected sub-graph $G'$ containing all tables in $\mathcal{T}'$ from graph $G$.
\textit{Steiner Tree Problem} is an NP-Hard problem. We employ a burst-search algorithm to get the solution thanks to the limited size of the schema currently. Incorporating new algorithms~\cite{DBLP:STP} for the larger database is left as future work.

For a high recall to avoid the error propagation problem, the table with the highest probability under $\tau_p$ will be included in graph $G'$ if the table has an edge with a node in $G'$. All nodes in $G'$ are denoted as $\mathcal{T}'$ for the kept tables.
For each table $t_i$ in $\mathcal{T}'$, columns with a probability exceeding $\tau_p$ and the \textit{primary keys} are kept, denoted as $\mathcal{C}_{i}'$. We define $\tau_n$ as the minimum column number to keep the table semantics.

Following the pruning module, only the target SQL-relevant schema information remains. Any primary keys in $\mathcal{P}$ and foreign keys in $\mathcal{F}$ that are unrelated to the tables $\mathcal{T}'$ and columns $\mathcal{C}'$ will be discarded.  For consistency and ease of notation, we continue to denote the pruned schema as $\mathcal{D}=<\mathcal{T}, \mathcal{C}, \mathcal{P}, \mathcal{F}>$.

\begin{figure}
  \centering
  \includegraphics[width=\linewidth]{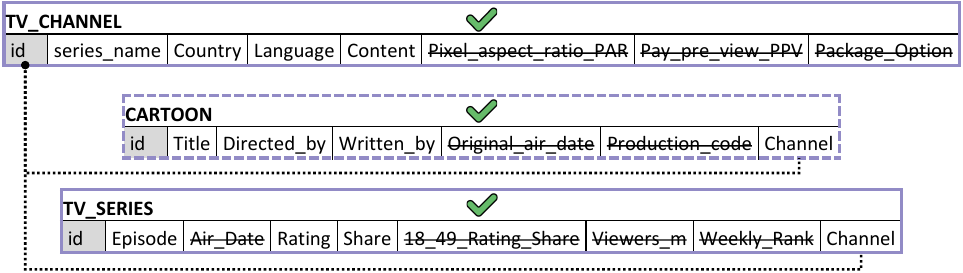}
  \caption{Schema pruning.}
  \label{fig:schema_pruning}
  \vspace{-3mm}
\end{figure}

For the example illustrated in Figure~\ref{fig:intro_example}, a trained classifier calculates the relevance of each table and column to the NL query. As shown in Figure~\ref{fig:schema_pruning}, tables with a probability exceeding $\tau_p$ are outlined with a solid purple line. Table \textit{CARTOON} has the highest probability among tables with probabilities below $\tau_p$, marked by a dashed purple line. We include table \textit{CARTOON} for high recall. The columns with a strikethrough will be removed when we set $\tau_n = 5$.

PURPLE focuses on keeping tables that are connected, improving efficiency. 
It also includes tables likely to be misclassified to boost recall without much extra cost.

\subsection{Skeleton Prediction}\label{sec:skeleton_prediction}

\begin{figure}
  \centering
  \includegraphics[width=\linewidth]{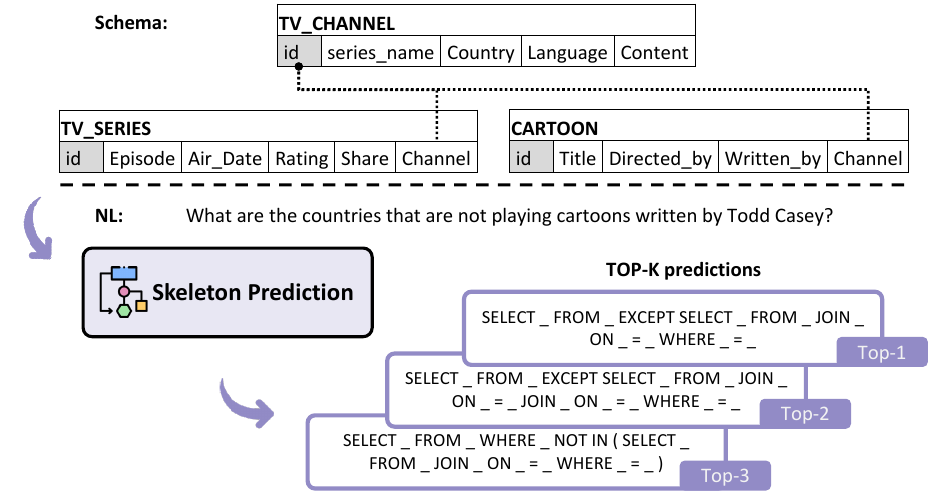}
  \caption{Skeleton prediction.}
  \label{fig:skeleton_prediction}
  \vspace{-3mm}
\end{figure}

Detecting the requisite operator composition is crucial for acquiring the necessary knowledge for LLMs. We notice that existing PLMs-based approaches achieve high Exact Match accuracy, suggesting that fine-tuning enables PLMs to identify operator compositions. Moreover, we propose a PLMs-based skeleton prediction module. The module uses the top-$k$ predicted skeletons, which have more operator composition diversity than predicted SQL queries. This strategy ensures a high recall of the requisite operator compositions, recognizing that the predicted skeleton is an intermediary rather than the terminal output compared with the PLMs-based approaches.

Our skeleton generator is built on sequence-to-sequence PLMs. The training loss function can be formulated as:
$$\mathcal{L}_{gen} = -\sum_{(\mathcal{X}, \mathcal{D}, \mathcal{S}) \in Train} \sum_{i=1}^{|\mathcal{S}|} \log P(\mathcal{S}_{i}|\mathcal{S}_{< i},\mathcal{X},\mathcal{D})$$
We process the gold SQL to obtain the target skeleton $\mathcal{S}$ for each training data. Every database-specific entity, including tables, columns, values, and aliases, is replaced by underscores.

We obtain the top-$k$ outputs using beam search~\cite{DBLP:beam_search}.
At step $i$, the skeleton token $\mathcal{S}_i$ is determined by:
$$\mathcal{S}_i=\mathop{\arg \max}_{v \in V}P(v|\mathcal{S}_{< i},\mathcal{X}, \mathcal{D})$$
$V$ represents the vocabulary of the PLM. The beam search halts upon encountering the stop token. For each $\mathcal{S}$ output, its sequence probability is computed as:
$$P(\mathcal{S}) = \prod \limits_{i=1}^{|\mathcal{S}|}P(\mathcal{S}_i|\mathcal{S}_{< i},\mathcal{X}, \mathcal{D})$$

We choose T5~\cite{DBLP:T5} as the PLM for skeleton prediction. As illustrated in Figure~\ref{fig:skeleton_prediction}, the top-$3$ predicted skeletons for the task in Figure~\ref{fig:intro_example} are presented.
The target skeleton is predicted by the skeleton model as the top-1 output.

The specialized skeleton prediction model has the ability to distinguish requisite composition knowledge due to fine-tuning. This model offers two primary advantages over the PLMs-based NL2SQL models.
Firstly, skeleton generation abstracts away from SQL details, simplifying the complexity of the task. 
Secondly, the top-$k$ predictions generated by this model exhibit a higher degree of diversity because the same skeleton with different database-specific tokens are ignored.

\subsection{Demonstration Selection}\label{sec:demonstration_selection}

\begin{figure}
  \centering
  \includegraphics[width=\linewidth]{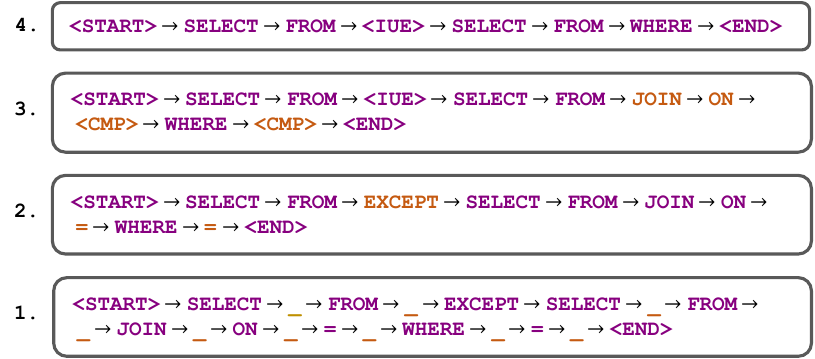}
  \caption{Automaton abstraction example.}
  \label{fig:automaton_abstraction}
  \vspace{-3mm}
\end{figure}

The main idea of PURPLE is to select a set of demonstrations that contains the necessary logical operator composition, thereby instructing LLMs on generating accurate SQL queries. However, this demonstration-based approach has several challenges when relying on predicted SQL skeletons:

\begin{itemize}
    \item A selection of demonstrations that precisely match the predicted skeletons will introduce the generalization problem. Given the infinite logical compositions, a finite set of demonstrations will not be enough for all tasks.
    
    \item Skeleton prediction accuracy depends on the PLMs used. Even with a top-$k$ strategy for skeleton prediction, achieving complete recall of the target skeleton remains hard. Therefore, enhancing the selection process for potential inaccuracies in the predicted skeleton is important.

    \item SQL is a complex declarative language that presents difficulties in capturing logical operator composition similarity.
    Ineffective similarity measures can introduce noise,
    failing to teach the LLMs to handle the NL2SQL task.

\end{itemize}

To capture the logical composition knowledge inherent in the demonstrations and to overcome the challenges mentioned above, we propose an automaton-based modeling of SQL composition knowledge with a four-level abstraction hierarchy. An automaton, characterized as a sequence of states, represents a strict operator composition structure. We introduce four-level abstractions to enhance this automaton with generalization and fuzzification capacity. This hierarchical automaton modeling design enables PURPLE to discern the logical operator composition and extract pertinent demonstrations.

\subsubsection{Automaton Modeling}

\begin{figure}
  \centering
  \includegraphics[width=\linewidth]{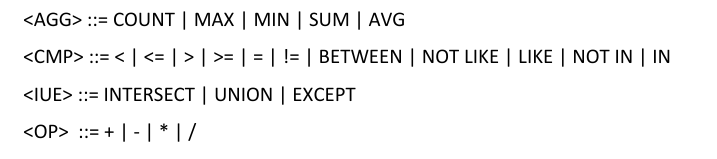}
  \caption{Structure-Level abstraction mapping rules.}
  \label{fig:structure_level_automaton}
  \vspace{-3mm}
\end{figure}

The sequence of SQL operators, comprising various keywords and their order, conveys distinct semantic compositions. We conceptualize this logical operator composition through a hierarchical abstraction within an automaton framework, thereby encapsulating compositional knowledge across varying granularity. The four discrete abstraction levels of this automaton are named \textbf{Detail-Level}, \textbf{Keywords-Level}, \textbf{Structure-Level}, and \textbf{Clause-Level}, each representing a more coarse-grained composition of the SQL query. Figure~\ref{fig:automaton_abstraction} illustrates an automation abstraction example of the skeleton shown in Section~\ref{subsec:sql_skeleton}. A detailed description of these levels is as follows:

\renewcommand{\labelenumi}{\theenumi.}
\begin{enumerate}
    
    \item \textbf{Detail-Level}: This level captures each component based on the predicted skeleton. It preserves the placeholders for columns and tables, reflecting the quantity and position of database-related elements.
    
    \item \textbf{Keywords-Level}: This level abstracts the placeholders to concentrate on SQL keywords. It contains all SQL keywords to reflect the logical operator composition, such as the comparison operator ``{\fontfamily{pcr}\selectfont \textcolor{darkpurple}{\textbf{=}}}''. This abstraction level shifts the focus solely to the logical operators within SQL.
    
    \item \textbf{Structure-Level}: Specific logical operators are classified under broader categories. For instance, ``{\fontfamily{pcr}\selectfont \textcolor{darkpurple}{\textbf{=}}}'' is generalized to ``{\fontfamily{pcr}\selectfont \textcolor{darkpurple}{\textbf{<CMP>}}}'', and ``{\fontfamily{pcr}\selectfont \textcolor{darkpurple}{\textbf{EXCEPT}}}'' to ``{\fontfamily{pcr}\selectfont \textcolor{darkpurple}{\textbf{<IUE>}}}''. This abstraction masks the detailed semantics, enabling the automaton to capture the structural semantics. The mapping rules of this level are shown in Figure~\ref{fig:structure_level_automaton}.

    \item \textbf{Clause-Level}: Representing the highest level of abstraction, this level concentrates on the principal clauses of the SQL query, masking all details within those clauses. Operators like ``{\fontfamily{pcr}\selectfont \textcolor{darkpurple}{\textbf{WHERE}}}'' and ``{\fontfamily{pcr}\selectfont \textcolor{darkpurple}{\textbf{<IUE>}}}'' are kept for the clause level semantics.
    
\end{enumerate}

Previous studies, such as DAIL-SQL~\cite{DBLP:DAIL-SQL}, mainly focus on the Keywords-Level similarity. However, they typically overlook the keyword order because they rely on Jaccard Similarity calculations. In contrast, PURPLE models logical composition through a four-level automaton, representing keyword selection and ordering. This method of demonstration selection via the automaton framework facilitates the selection of essential composition knowledge by LLMs, which is not addressed by previous research.

For instance, DAIL-SQL~\cite{DBLP:DAIL-SQL} considers a skeleton like: 

{\footnotesize \fontfamily{pcr}\selectfont \textcolor{darkpurple}{\textbf{SELECT}}\;\_\;\textcolor{darkpurple}{\textbf{FROM}}\;\_\;\textcolor{darkpurple}{\textbf{JOIN}}\;\_\;\textcolor{darkpurple}{\textbf{ON}}\;\_\;\textcolor{darkpurple}{\textbf{=}}\;\_\;\textcolor{darkpurple}{\textbf{WHERE}}\;\_\;\textcolor{darkpurple}{\textbf{=}}\;\_}

{\footnotesize \fontfamily{pcr}\selectfont \textcolor{darkpurple}{\textbf{EXCEPT}}}

{\footnotesize \fontfamily{pcr}\selectfont \textcolor{darkpurple}{\textbf{SELECT}}\;\_\;\textcolor{darkpurple}{\textbf{FROM}}\;\_}

\noindent as same with the skeleton shown in Section~\ref{subsec:sql_skeleton}. However, it failed to provide accurate composition knowledge for LLMs. Conversely, PURPLE prioritizes demonstrations as exemplified in Figure~\ref{fig:demonstration}, as these can be matched through the Structure-Level automaton. The automaton design enhances the ability of PURPLE to select more relevant compositional knowledge for LLMs, improving overall effectiveness in SQL writing.

\subsubsection{Automaton Construction}\label{subsubsec:automaton_construction}

The automaton is constructed by parsing SQL skeletons extracted from all of the demonstrations $\mathcal{E}$. For each demonstration $e_i$ as shown in Section~\ref{subsec:demonstrations}, we mask the database-specific tokens in the target SQL $\mathcal{Y}^{e_i}$ to get the skeleton $\mathcal{S}^{e_i}$. We parse all skeletons into basic elements, that we use to construct the Detail-Level automaton. In addition, we add two specialized state nodes, denoted as ``{\small \fontfamily{pcr}\selectfont \textcolor{darkpurple}{\textbf{<START>}}}'' and ``{\small \fontfamily{pcr}\selectfont \textcolor{darkpurple}{\textbf{<END>}}}'', which serve as the initial and terminal states respectively. As the level of abstraction increases, more details are progressively masked.

We build the automaton for demonstration selection. To accelerate the selection process, we store the index of each demonstration within the ``{\small \fontfamily{pcr}\selectfont \textcolor{darkpurple}{\textbf{<END>}}}'' state node of its corresponding automaton. 
As we process the predicted skeleton through to the ``{\small \fontfamily{pcr}\selectfont \textcolor{darkpurple}{\textbf{<END>}}}'' state, the stored index helps to retrieve all demonstrations sharing identical automaton states. 
An \textit{empty list} will be returned if a state sequence is absent in the demonstrations.
Furthermore, we will remove all of the out-of-vocabulary tokens before parsing the predicted skeletons, which are introduced by the skeleton prediction model.

\subsubsection{Automaton Matching}\label{subsubsec:automaton_matching}

Our approach takes only identical sequences of automaton states as matches. This approach simplifies the extraction of demonstrations that align with each predicted skeleton across four levels of abstraction. Leveraging both the top-predicted skeletons and multiple abstraction levels, selecting demonstrations is a non-trivial task.

\begin{algorithm}
\small
 \SetAlgoLined
 \SetKwFunction{matchAutomaton}{\textsc{Match}}
 \SetKwFunction{removeEmpty}{\textsc{Remove}-Empty}
 \SetKwFunction{notEmpty}{\textsc{Not}-Empty}
 \SetKwFunction{getTop}{\textsc{Get}-Top}
 \SetKwFunction{selectDemonstration}{\textsc{Pop}-Demo}
 \SetKwFunction{increaseGeneralization}{\textsc{Increase}-Generalization}
 \SetKwFunction{demonstrationRanking}{\textsc{Demonstration}-Selection}
 
 \SetKwInOut{KwIn}{Inputs}
 \SetKwInOut{KwOut}{Output}
 \KwIn{\small{Automaton list $\mathcal{A}$}; Query instance $\mathcal{Q}$}
 \KwOut{\small{Selected demonstrations $\mathcal{E}'$}}
\SetKwProg{Fn}{Procedure}{:}{}
 \Fn{\demonstrationRanking{$\mathcal{A}$, $\mathcal{Q}$}}{
    $\mathcal{I}, \mathcal{E}' \gets \texttt{[]}; p \gets p_0$
    
    \For{\textup{\textbf{each} $i \in \texttt{[}1, ..., 4\texttt{]}$}}{
        \For{\textup{\textbf{each} $j \in \texttt{[}1, ..., k\texttt{]}$}}{
            \tcp{\color{darkblue}Get index by automaton}

            $\mathcal{I}$.append(\matchAutomaton{$\mathcal{A}$[$i$], $\mathcal{Q}$.pred[$j$]})

        }
    }

    \While{\notEmpty{$\mathcal{I}$}}{
        
        \For{\textup{\textbf{each} $a \in $ \getTop{$\mathcal{I}$, $p$}}}{

        \tcp{\color{darkblue} Select demonstrations}

        $\mathcal{E}'$.append(\selectDemonstration{$a$, $\mathcal{E}'$})

        }

        \tcp{\color{darkblue} Higher generalization ability}
        
        $p$ $\gets$ \increaseGeneralization{$p$}
        
    }

    \KwRet $\mathcal{E}'$
  }
 \caption{Demonstration Selection Algorithm}
 \label{algoritm:demonstration_selection}
\end{algorithm}

PURPLE gives preference to skeletons that have high probability according to the model predictions and correspond to matches at lower levels of abstraction. This is based on the understanding that a higher probability prediction coupled with a lower abstraction level typically indicates a more precise match. Conversely, lower predicted probabilities and matches at higher levels of abstraction indicate greater generalization capacity but introduce more noise. PURPLE tries to balance the robustness and efficiency of the selection process.

The demonstration selection algorithm is shown in Algorithm~\ref{algoritm:demonstration_selection}, which prioritizes the selection of demonstrations with a higher prediction probability and a lower level of abstraction.
The input $\mathcal{A}$ is the constructed automaton as shown in Section~\ref{subsubsec:automaton_construction}, and $\mathcal{A}[i]$ represents the automaton at abstraction level $i$. The query instance $\mathcal{Q}$ stores the predicted skeletons as $\mathcal{Q}.pred$, and $\mathcal{Q}.pred[j]$ represents the $j$-th skeleton.
The {\fontfamily{pcr}\selectfont \textsc{Match}} function (line 5) identifies indices of demonstrations that align with the $j$-th skeleton at abstraction level $i$. The preferential matching sequence $\mathcal{I}$ is a list with a size of $4*k$, which stores matched indices (lines 2-5).
The parameter $p$, for balancing precision and generalization, starts at $p_0$ and is adjusted by {\fontfamily{pcr}\selectfont \textsc{Increase-Generalization}}. As $p$ increases, more demonstrations are considered (line 7). The  {\fontfamily{pcr}\selectfont \textsc{Get-Top}} function selects the top-$p$ indices, while  {\fontfamily{pcr}\selectfont \textsc{Pop-Demo}} retrieves matching demonstrations, ensuring compatibility across abstraction levels and avoiding duplicates in $\mathcal{E}'$.

\begin{figure}
  \centering
  \includegraphics[width=\linewidth]{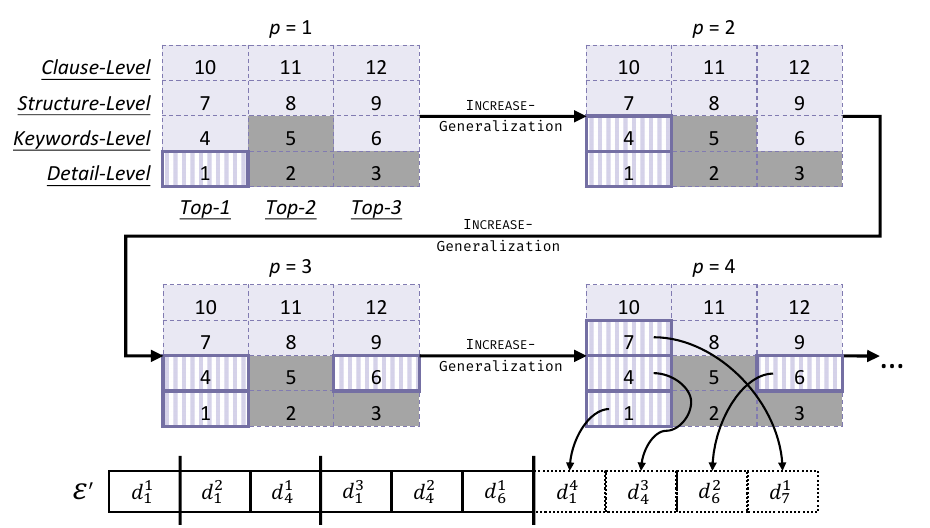}
  \caption{Demonstration selection example.}
  \label{fig:demonstration_selection_example}
  \vspace{-3mm}
\end{figure}

Taking Figure~\ref{fig:demonstration_selection_example} as an example, we represent the $\mathcal{I}$ as a matrix and discuss the detailed selection process (lines 6-9).
In this matrix, columns correspond to the top-$k$ (with $k=3$ in our example) predicted skeletons, and rows correspond to four abstraction levels. Gray cells within the matrix indicate the absence of matching demonstrations for that specific combination. For instance, cell 2 means missing the Detail-Level match for the second skeleton. We start with $p=1$ and increase it by 1 at each iteration. We select the top-$p$ matches at each step, highlighted by purple strips in the figure. 
Each demonstration added to the selected demonstration queue $\mathcal{E}'$ is represented as $d_i^j$, meaning the $j$-th demonstration from the $i$-th cell. For example, in the first step with $p=1$, demonstration $d_1^1$ is added to $\mathcal{E}'$, and in the second step with $p=2$, $d_1^2$ and $d_4^1$ are added, as cells $2$ and $3$ lack matches. This process continues until no further demonstrations are contained in $\mathcal{I}$.

The value of $p_0$ and the {\fontfamily{pcr}\selectfont \textsc{Increase}-Generalization} function could be guided by the size of the automaton. A smaller automaton size suggests a higher density of demonstrations within each automaton state, which may introduce greater noise into the selection process. For instance, in our analysis of the Spider benchmark, we analyze the distribution of ``{\small \fontfamily{pcr}\selectfont \textcolor{darkpurple}{\textbf{<END>}}}'' states and their respective distribution within the four levels of automaton abstraction, finding proportions of $912:708:363:59$. Consequently, we set $p_0$ to $1$, with $p$ increasing by $1$ at every step, aiming for a simplified expected matching ratio of $4:3:2:1$ across the abstraction levels, balancing precision and generalization.
The remaining demonstrations are chosen randomly to fully utilize the budget.

Automaton with four-level abstraction can model logical operator composition knowledge across varying granularities, which is advanced in augmenting both the generalization and fuzzification capacities for the demonstration selection. We acknowledge that matching demonstrations at a higher level of abstraction can encompass broader logic by masking finer details, which could introduce uncertainty into the selection process. However, this broader perspective is crucial because it retains the fundamental knowledge of logical operator compositions, and the introduced minor errors could be identified and fixed by LLMs.
Through this approach, PURPLE could extract demonstrations that contain the requisite operator composition knowledge, thereby enhancing its performance.

\subsection{Database Adaption}\label{sec:databse_adaption}

The hallucination problem of existing LLMs results in the generation of invalid SQL during NL2SQL translation. Especially, SQL is related to the database schema and DBMS. Such a problem will cause a performance decline and lead to inconsistent translations. Through a detailed analysis of the LLMs outputs, we categorize common errors and develop algorithms to adapt the generated SQL to specific database schema and SQL dialect. We also incorporate an execution-consistency strategy to stabilize the translation outputs.

\subsubsection{SQL Adaption}

Modern LLMs benefit from extensive pre-training corpora, which equips the model with basic SQL knowledge. However, the corpora contain SQL variations from multiple DBMSs,
which can result in translations that include syntax not supported by current DBMS.

\begin{table}[h]
  \centering
  \caption{Error types and corresponding example}
  \begin{adjustbox}{width=1.0\linewidth}
   \begin{tabular}{c | l} \Xhline{3\arrayrulewidth}
     \textbf{Error Type} & \textbf{Example}  \\
     \hline

     \makecell[c]{Table-Column-Mismatch} 
     &
     \makecell[l]{
     {\fontfamily{pcr}\selectfont SELECT \textcolor{darkred}{\textbf{T2.title}} FROM cartoon AS T1 JOIN} \\ 
     {\fontfamily{pcr}\selectfont  tv\_channel AS T2 ON T1.channel = T2.id }
      \\
     {\fontfamily{pcr}\selectfont WHERE T2.series\_name = "Sky Radio";}
     } \\
     \hline 
     
     \makecell[c]{Column-Ambiguity} 
     &
     \makecell[l]{
     {\fontfamily{pcr}\selectfont SELECT \textcolor{darkred}{\textbf{maker}}, model FROM car\_makers} \\ 
     {\fontfamily{pcr}\selectfont JOIN model\_list ON car\_makers.id = } \\
     {\fontfamily{pcr}\selectfont model\_list.maker JOIN car\_names } \\
     {\fontfamily{pcr}\selectfont  ON model\_list.model = car\_names.makeid;}
     } \\
     \hline 

     \makecell[c]{Missing-Table} 
     &
     \makecell[l]{
     {\fontfamily{pcr}\selectfont SELECT COUNT(DISTINCT language)} \\ 
     {\fontfamily{pcr}\selectfont FROM countrylanguage} \\ 
     {\fontfamily{pcr}\selectfont WHERE isofficial = 'T' AND \textcolor{darkred}{\textbf{indepyear}} < 1930;}
     } \\
     \hline 
     
     \makecell[c]{Function-Hallucinations} 
     &
     \makecell[l]{
     {\fontfamily{pcr}\selectfont SELECT \textcolor{darkred}{\textbf{CONCAT(first\_name, ' ', last\_name)}}} \\ 
     {\fontfamily{pcr}\selectfont AS full\_name FROM players} \\
     {\fontfamily{pcr}\selectfont ORDER BY birth\_date;}
     } \\
     \hline 
          
     \makecell[c]{Schema-Hallucinations} 
     &
     \makecell[l]{
     {\fontfamily{pcr}\selectfont SELECT \textcolor{darkred}{\textbf{T1.course\_id}}, COUNT(*)} \\ 
     {\fontfamily{pcr}\selectfont AS count FROM transcript\_contents AS T1} \\
     {\fontfamily{pcr}\selectfont JOIN student\_enrolment\_courses AS T2 ON} \\
     {\fontfamily{pcr}\selectfont T1.student\_course\_id = T2.student\_course\_id} \\
     {\fontfamily{pcr}\selectfont JOIN transcripts AS T3 ON } \\
     {\fontfamily{pcr}\selectfont T1.transcript\_id = T3.transcript\_id } \\
     {\fontfamily{pcr}\selectfont GROUP BY T1.course\_id ORDER BY count DESC;} \\
     } \\
     \hline 

     \makecell[c]{Aggregation-Hallucinations} 
     &
     \makecell[l]{
     {\fontfamily{pcr}\selectfont SELECT \textcolor{darkred}{\textbf{COUNT(DISTINCT series\_name, content)}}} \\
     {\fontfamily{pcr}\selectfont FROM tv\_channel;}
     } \\
     \hline

     \Xhline{3\arrayrulewidth}
   \end{tabular}
  \end{adjustbox}
  \label{tab:error_types}
 \end{table}

We identify six primary error categories, as shown in Table~\ref{tab:error_types}, which provides examples of invalid SQL for each category. We design heuristic algorithms for each error type. The algorithms fix the SQL queries that result in execution errors. So that PURPLE ensures that the SQL adaption strategy does not introduce undesired side effects to the valid SQL.

\begin{itemize}

    \item \textbf{Table-Column Mismatch:} LLMs reason based on statistical patterns,
    leading to the incorrect alignment of columns to tables. As illustrated in Table~\ref{tab:error_types}, the column \textit{title} belongs to the table \textit{cartoon}, rendering \textit{T2.title} an error. We rectify such errors by mapping the column to its correct table and adjusting the table identifier accordingly.

    \item \textbf{Column-Ambiguity:} A SQL might be invalid if multiple tables contain a column with the same name ambiguity. We randomly assign the column to one of its potential tables, ensuring its unique identification.

    \item \textbf{Missing-Table:} As denoted in Table~\ref{tab:error_types}, the column \textit{indepyear} belongs to the table \textit{country}, which is absent in the SQL. We fix this by including the table into the FROM clause based on primary-foreign key relationships.
    
    \item \textbf{Function-Hallucinations:} Certain functions like \textit{CONCAT} are not supported in SQLite, resulting in invalid SQL. Our immediate solution is to omit the unsupported function call. An optimal solution would involve mapping functions across different DBMSs for future work.

    \item \textbf{Schema-Hallucinations:} LLMs may generate SQL referencing non-existent tables or columns within a given schema. For instance, Table~\ref{tab:error_types} highlights that the column \textit{course\_id} is not present in any of the tables. We tackle this by identifying and substituting it with a column having a minimal string edit distance.
    
    \item \textbf{Aggregation-Hallucinations:} Aggregation functions in SQLite are designed to take a single column as input. To rectify errors like the one in Table~\ref{tab:error_types}, we divide the \textit{COUNT} function into two separate counts, preserving the \textit{DISTINCT} keyword for both columns.

\end{itemize}

PURPLE offers solutions for the six most common LLM-induced SQL errors. In our implementation, we attempt to rectify a non-executable SQL up to five times.

\subsubsection{Consistency Strategy}

Existing works like SQL-PaLM~\cite{DBLP:SQL-PALM}, C3~\cite{DBLP:C3}, and DAIL-SQL~\cite{DBLP:DAIL-SQL} integrate the execution-consistency strategy in stabilizing LLMs-based NL2SQL translations. We integrate this strategy into PURPLE with an increase in the cost of output tokens.

In detail, PURPLE prompts the LLMs to produce $n$ SQL translations for every API call. SQL adaption process will be executed for the generated invalid SQL. Subsequently, each executable SQL is executed against the database. PURPLE then employs a voting mechanism based on the SQL execution results. The first SQL that yields the consensus execution result is selected as the output.

The hallucination of LLMs is an unavoidable issue. A categorization of issues stemming from hallucinations is beneficial in fixing those bugs. The fixing process is safe because it does not have side effects on the executable SQL.
The database adaptation module utilizes database insights and DBMS characteristics to efficiently rectify erroneous SQL queries.

\section{Experiments}\label{sec:experiments}

In this section, we evaluate the overall performance of PURPLE.
We discuss the trade-off between performance and API cost.
Furthermore, we explore the robustness of PURPLE  with various hyper-parameters and LLMs.
Additionally, we conduct ablation studies on each module.

\subsection{Experimental Setup}

\subsubsection{Benchmarks}

We evaluate PURPLE on four popular NL2SQL benchmarks: Spider~\cite{DBLP:Spider}, Spider-DK~\cite{DBLP:SPIDER-DK}, Spider-SYN~\cite{DBLP:SPIDER-SYN}, and Spider-Realistic~\cite{DBLP:SPIDER-REALISTIC}. 
The statistics about these benchmarks can be found in Table~\ref{tab:benchmark_statistics}.

\begin{table}
  \centering
  \caption{The statistics of NL2SQL benchmarks}
  \begin{adjustbox}{width=1.0\linewidth}
  \begin{tabular}{|c|c|c|c|c|} \hline
    \textbf{Benchmark} & \textbf{Queries} & \textbf{Databases} & \textbf{\makecell{Average length \\of NL queries}} & \textbf{\makecell{Average length \\of target SQL}}
    \\ \hline\hline
    \textsc{Spider(Train)} & 8,659 & 146 & 66.6  & 122.9  \\  \hline
    \textsc{Spider(Validation)} & 1,034 & 20 & 68.0  & 106.7  \\  \hline
    \textsc{Spider-DK} & 535 & 10 & 66.0 & 109.5  \\  \hline
    \textsc{Spider-Realistic} & 508 & 20 & 64.8  & 115.3  \\  \hline
    \textsc{Spider-SYN} & 1,034 & 20 & 68.8  & 106.7  \\  \hline
    
  \end{tabular}
  \end{adjustbox}
  \label{tab:benchmark_statistics}
  \vspace{-3mm}
\end{table}

\textbf{Spider} is a popular benchmark for NL2SQL translation, consisting of 200 databases with multiple tables and 10,181 NL-to-SQL pairs. It demands a comprehensive understanding of multi-table database relations, targeting performance evaluation on complex SQL translation based on unfamiliar domains.
We evaluate PURPLE on the validation set of Spider, and we take the training set as the demonstration.

\textbf{Spider-DK} is a more challenging version of the Spider validation set. Such a benchmark requires the NL2SQL strategy to know about domain-specific knowledge for the SQL generation. Preliminary observations indicate many approaches struggle with this heightened domain-specific demand.

\textbf{Spider-Realistic} emphasizes the challenges of text-table alignments. It provides a more realistic scenario by omitting explicit mentions of column names and requires approaches to map NL terms to relevant database schema items adeptly.

\textbf{Spider-SYN} stems from the Spider. It modifies NL queries by swapping schema-related terms with handpicked synonyms, challenging the reliance on lexical matching.

\subsubsection{Evaluation Metrics}

We employ three evaluation metrics to assess the performance of PURPLE comprehensively: \textit{Exact-Set Match (EM)} accuracy, \textit{Execution Match (EX)} accuracy~\cite{DBLP:Spider}, and \textit{Test-Suite (TS)} accuracy~\cite{DBLP:TEST-SUITE}.
The detailed description of the metrics is as follows.

\textbf{EM} accuracy is one of the official evaluation metrics of Spider, which uses a set comparison for each clause. 
While precise, EM might yield false negatives due to new syntax structures from semantic parsers.

\textbf{EX} accuracy is also officially supported by Spider, which checks the congruence of executed predicted SQL query results with expected outcomes. EX can sometimes return false positives when differing SQL queries yield identical results but potentially with varied semantics.

\textbf{TS} accuracy aims to rectify the EX metric by employing a distilled test suite of databases~\cite{DBLP:TEST-SUITE}. The distilled database is created by selecting a small subset from numerous random databases that can distinguish between correct and nearly correct queries, ensuring high code coverage. We follow the original script\footnote{\url{https://github.com/ruiqi-zhong/TestSuiteEval}} to generate an augmented 100-fold distilled database for evaluation.

\subsubsection{Baselines} 

Existing NL2SQL approaches are used for comparison to show the performance of PURPLE. We choose some SOTA LLMs-based approaches for comparison, including ChatGPT-SQL~\cite{DBLP:ChatGPT-zero-shot}, C3~\cite{DBLP:C3}, DIN-SQL~\cite{DBLP:DIN-SQL} and DAIL-SQL~\cite{DBLP:DAIL-SQL}. Basic few-shot strategies are shown in DIN-SQL~\cite{DBLP:DIN-SQL}, and we also include the GPT4 results for comparison. We also report the performance of some PLMs-based approaches on Spider for reference, including PICARD~\cite{DBLP:PICARD}, RASAT~\cite{DBLP:RASAT}, RESDSQL~\cite{DBLP:RESD} and Graphix-T5~\cite{DBLP:GRAPHIX}.

\textbf{ChatGPT-SQL} aims to thoroughly assess the zero-shot NL2SQL capabilities of ChatGPT. The predictions from this approach have been made publicly available, and we leverage these open-source results for our comparative analysis.

\textbf{C3} is a zero-shot LLMs-based approach by hand-crafted instruction.  C3 also proposes to reduce the input length of LLM API calls but fails to control the output length.

\textbf{DIN-SQL} employs a few-shot approach and has achieved leading performance in terms of EX on the Spider. DIN-SQL incorporates CoT for performance enhancement. Additionally, DIN-SQL reports the result of \textbf{GPT4 few-shot} and \textbf{GPT4 zero-shot} approaches, which we include in our comparisons.

\textbf{DAIL-SQL} implement demonstration selection by analyzing NL query and SQL similarity. This adaptable demonstration selection strategy has shown promising results, especially when integrated with the capabilities of GPT4.

\textbf{PICARD}, \textbf{RASAT}, \textbf{RESDSQL}, and \textbf{Graphix-T5} represent SOTA PLMs-based methods. They are all based on the T5 model with improving the encoder, decoder, or task formulation. We report their optimal performance for comparison.

\subsubsection{Implementation Details}

We employ ChatGPT (gpt-3.5-turbo-0613)\footnote{\url{https://openai.com/chatgpt}} and GPT4 (gpt-4-0613)\footnote{\url{https://openai.com/gpt-4}} for the SQL generation. Our training environment operates on Centos 7.9, with a 64-core CPU, 512GB of memory, and 8 NVIDIA A100 GPUs. For the schema pruning module, we set $\tau_p = 0.5$, $\tau_n = 5$. We fine-tune a T5-3B model for skeleton prediction, selecting the top-$3$ skeletons. The automaton matching hyper-parameters followed the setting shown in Section~\ref{subsubsec:automaton_matching}. 
For the cost saving, comparisons involving GPT4-based approaches are confined to Section~\ref{subsec:overall_performance} and Section~\ref{various_llms}. All other experimental evaluations are conducted using ChatGPT.

\subsection{Overall Performance}\label{subsec:overall_performance}

We evaluate PURPLE by comparing it against SOTA LLMs-based and PLMs-based approaches for a comprehensive view.

\begin{table}[h]
    \setlength\tabcolsep{6pt}
    \centering
    \caption{Translation accuracy on Spider.}
    \begin{tabular}{l c c c}
        \hline
        \textbf{Strategy}        & \textbf{EM\%} & \textbf{EX\%} & \textbf{TS\%} \\ 
        \hline
        \hline
        PICARD        &     75.5 &     79.3 &     69.4  \\
        RASAT        &     75.3 &     80.5 &     70.3  \\
        RESDSQL   &     \textbf{80.5} &     84.1 &     73.5  \\
        Graphix-T5   &     77.1 &     81.0 &     74.9  \\
        \hline
        ChatGPT-SQL (ChatGPT)           &     37.9 &     70.1 &     60.1     \\
        C3 (ChatGPT)                    &     43.1 &     81.8 &     72.1     \\
        Zero-shot (GPT4)                &     42.4 &     72.9 &     64.9     \\
        Few-shot (GPT4)                 &     54.3 &     76.8 &     67.4     \\
        DIN-SQL (GPT4)                  &     60.1 &     82.8 &     74.2     \\
        DAIL-SQL (GPT4)                 &     68.7 &     83.6 &     76.2     \\
        \rowcolor{purple!28} PURPLE (ChatGPT)                &     76.1 &     84.8 &     80.1     \\
        \rowcolor{purple!28} PURPLE (GPT4)                   &     \textbf{80.5} &     \textbf{87.8} &     \textbf{83.3}  \\
        \hline
        \end{tabular}
    \label{tab:overall_performance}
\end{table}

Table~\ref{tab:overall_performance} illustrates that when augmented with GPT4, PURPLE surpasses other LLMs-based strategies across all metrics on the validation set of Spider, including EM, EX, and TS. Remarkably, PURPLE remains superior even when coupled with the comparatively weak ChatGPT. DAIL-SQL achieves an 83.6\% EX among its LLMs-based counterparts but only 76.2\% on TS. 
PURPLE with GPT4 enhances the performance by a large margin, which means a 4.2\% improvement on EX and a 7.1\% improvement on TS than DAIL-SQL. A typical challenge for existing LLMs-based NL2SQL approaches is their low EM because of their inability to guide the generative process of LLMs. However, PURPLE achieves an 11.8\% improvement over DAIL-SQL and a 20.4\% improvement over DIN-SQL in EM, showing the reliability of PURPLE.

In addition, we compare PURPLE with SOTA PLMs-based approaches on the Spider. 
While PURPLE incorporates a fine-tuning phase, its primary application is demonstration retrieval to enhance the LLMs. PURPLE reaches the top EM score compared with all of the PLMs-based approaches, in which PURPLE achieves 80.5\% EM on the spider validation set. Achieving the highest score in EM, EX, and TS, PURPLE shows the ability of LLMs-based NL2SQL approaches to outperform their PLMs-based counterparts.

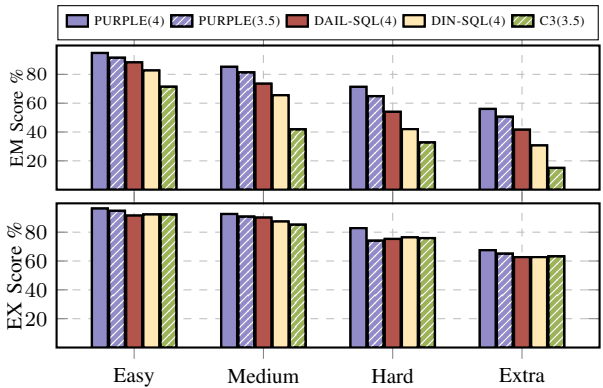
\begin{figure}[h]
    \centering
    \hspace{-3mm}
    \begin{tikzpicture}
    \begin{scope}        
    \begin{axis}[
        height=3.5cm, 
        width=8.8cm,
        ybar = .17mm, 
        enlarge y limits=0,
        enlarge x limits=0.2, 
        ymin = 0.0,
        ymax = 100.0,
        ylabel={EM Score \%},
        ylabel style={yshift=-2mm, font=\scriptsize},
        xlabel style={font=\scriptsize},
        axis line style=thick,
        symbolic x coords={Easy, Medium, Hard, Extra}, 
        xtick={},
        xtick=data,
        xticklabel style={font=\scriptsize,  text opacity=0},
        ytick={20,40,60,80},
        yticklabel style={font=\scriptsize},
        bar width=6pt,
        ymajorgrids=true,
        xmajorgrids=true,
        grid style=dashed,
        nodes near coords align={vertical}, 
        legend image code/.code={%
            \draw[#1, draw=black, thick] (0cm,-0.05cm) rectangle (0.28cm,0.1cm);
        },
        legend pos=north east,
        legend style={
            font=\tiny,
            at={(0.5,1.28)},
            legend columns=-1,
            anchor=north,
        }
    ]
        \addplot[
            line width= .3mm, 
            fill=purple,
            postaction={
                pattern color=white
            }
        ]
        coordinates {(Easy,94.8) (Medium,85.2) (Hard,71.3) (Extra,56.0)};
        \addplot[
            line width= .3mm,
            fill=purple,
            postaction={
                pattern=north east lines, 
                pattern color=white
            }
        ]
        coordinates {(Easy,91.5) (Medium,81.4) (Hard,64.9) (Extra,50.6)};
        \addplot [
            line width= .3mm,
            fill=cardinal,
            postaction={
                pattern color=white
             }
        ]
        coordinates {(Easy,88.3) (Medium,73.5) (Hard,54.0) (Extra,41.6)};
        \addplot [
            line width= .3mm,
            fill=tropicalrainforest,
            postaction={
                pattern color=white
             }
        ]
        coordinates {(Easy,82.7) (Medium,65.5) (Hard,42.0) (Extra,30.7)};
        \addplot [
            line width= .3mm,
            fill=palegreen,
            postaction={
                pattern=north east lines, 
                pattern color=white
             }
        ]
        coordinates {(Easy,71.4) (Medium,41.9) (Hard,32.8) (Extra,15.1)};
        \legend{\textsc{PURPLE(4)}, \textsc{PURPLE(3.5)}, \textsc{DAIL-SQL(4)}, \textsc{DIN-SQL(4)}, \textsc{C3(3.5)}}
    \end{axis}
    \end{scope}
    
    \begin{scope}[yshift=-2.1cm]
    \begin{axis}[
        height=3.5cm, 
        width=8.8cm,
        ybar = .17mm, 
        enlarge y limits=0,
        enlarge x limits=0.2, 
        ymin = 0.0,
        ymax = 100,
        ylabel={EX Score \%},
        ylabel style={yshift=-2mm, font=\footnotesize},
        axis line style=thick,
        symbolic x coords={Easy, Medium, Hard, Extra}, 
        xtick={},
        ytick={20,40,60,80},
        xtick=data,
        xticklabel style={font=\footnotesize},
        yticklabel style={font=\footnotesize},
        legend image code/.code={%
            \draw[#1, draw=black, thick] (0cm,-0.1cm) rectangle (0.2cm,0.12cm);
        },
        bar width=6pt,
        ymajorgrids=true,
        xmajorgrids=true,
        grid style=dashed,
        nodes near coords align={vertical}, 
        legend pos=north east,
        legend style={
            font=\tiny,
            at={(0.5,0.95)},
            legend columns=-1,
            anchor=north
        }
    ]
        \addplot[
            line width= .3mm,
            fill=purple,
            postaction={
                pattern color=white
            }
        ]
        coordinates {(Easy,96.4) (Medium,92.6) (Hard,82.8) (Extra,67.5)};
        \addplot[
            line width= .3mm,
            fill=purple,
            postaction={
                pattern=north east lines, 
                pattern color=white
            }
        ]
        coordinates {(Easy,94.8) (Medium,90.8) (Hard,74.1) (Extra,65.1)};
        \addplot [
            line width= .3mm,
            fill=cardinal,
            postaction={
                pattern color=white
             }
        ]
        coordinates {(Easy,91.5) (Medium,90.1) (Hard,75.3) (Extra,62.7)};

        \addplot [
            line width= .3mm,
            fill=tropicalrainforest,
            postaction={
                pattern color=white
             }
        ]
        coordinates {(Easy,92.3) (Medium,87.4) (Hard,76.4) (Extra,62.7)};
        
        \addplot [
            line width= .3mm,
            fill=palegreen,
            postaction={
                pattern=north east lines, 
                pattern color=white
             }
        ]
        coordinates {(Easy,92.3) (Medium,85.2) (Hard,75.9) (Extra,63.3)};
        
    \end{axis}
    \end{scope}
    \end{tikzpicture}
    \caption{Comparison of the EM/EX score on the Spider validation set regarding SQL hardness levels.}
    \label{fig:breakdown_analysis}
\end{figure}

Figure~\ref{fig:breakdown_analysis} shows the performance of various approaches based on SQL hardness levels for the Spider validation set. We follow the official evaluation scripts for the hardness classification. The legend shows the name of approaches and the LLMs, such as \textit{PURPLE(4)} means PURPLE with GPT4, \textit{C3(3.5)} represents C3 with ChatGPT. When augmented with GPT4, PURPLE consistently achieves the highest performance across all SQL hardness levels. Notably, even with the relatively weak ChatGPT, PURPLE still surpasses other approaches in terms of EM, regardless of SQL hardness. 

An observation is that PURPLE advances in handling the \textit{extra hard} SQL translations. This ability can be attributed to its emphasis on operator composition knowledge,
thereby enhancing the LLM with complex SQL generation capacity. 
Conversely, DIN-SQL employs CoT to facilitate LLMs in managing complex SQL constructions. While these CoT demonstrations help LLMs understand user intention, they fail to include SQL operator composition knowledge. 
In addition, C3 focuses on syntactic constraints within its designed instruction. 
However, such hand-crafted instructions are insufficient to provide the necessary operator composition knowledge. 
DAIL-SQL utilizes both NL and SQL similarity for demonstration selection, but the similarity function can not capture the operator composition similarity between two SQL queries as described in Section~\ref{subsubsec:automaton_construction}, thereby failing to address the limitations of LLMs. 
PURPLE selects the demonstrations based on the logical operator composition, which successfully improves the performance of existing general LLMs.

\subsection{Generalization Ability}

Generalization ability is a vital aspect when evaluating NL2SQL approaches. An NL2SQL system will likely be deployed on databases unseen during training. We utilize Spider-DK, Spider-SYN, and Spider-Realistic benchmarks to evaluate the generalization ability. We train PURPLE on the Spider dataset and test its EM and EX accuracy on the three benchmarks.
We compare the performance of two other ChatGPT-based strategies, ChatGPT-SQL and C3.

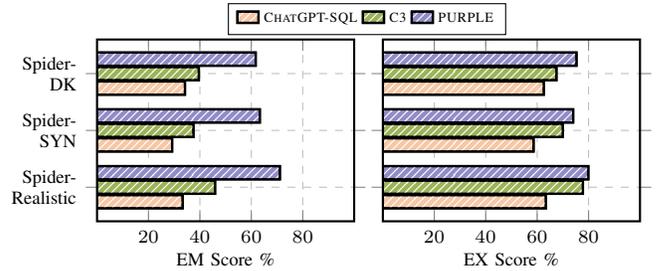
\begin{figure}[ht]
    \centering
    \vspace{-1mm}
    \begin{tikzpicture}
    \begin{scope}
        \begin{axis}[
        xbar = .17mm, 
        height=4cm, 
        width=5cm,
            xmin = 0,
            xmax = 100,
            xtick={20,40,60,80},
            enlarge y limits=0.3, 
            axis line style=thick,
            xlabel={EM Score \%},
            xlabel style={font=\scriptsize, yshift=1mm}, 
            xticklabel style={font=\scriptsize},
            symbolic y coords={Spider-Realistic, Spider-SYN, Spider-DK}, 
            yticklabel style={
        font=\scriptsize,
        text width=1cm, 
        align=right
    },
            ytick={Spider-Realistic, Spider-SYN, Spider-DK},
            bar width=5pt,
            xmajorgrids=true,
            ymajorgrids=true,
            grid style=dashed,
            nodes near coords align={vertical},
            legend image code/.code={
                \draw[#1, draw=black] (0cm,-0.05cm) rectangle (0.28cm,0.1cm);
            }, 
            legend style={font=\scriptsize, at={(1.05,1.2)},
                legend columns=-1,
                anchor=north,
                nodes={scale=0.75, transform shape}, 
            }
        ]

        \addplot [
            line width= .3mm, 
            fill=pink,
            postaction={
                pattern=north east lines, 
                pattern color=white
            }
        ]
        coordinates {(33.3,Spider-Realistic) (29.2,Spider-SYN) (34.2,Spider-DK)};
        
        \addplot [
            line width= .3mm, 
            fill=palegreen,
            postaction={
                pattern=north east lines, 
                pattern color=white
            }
        ]
        coordinates {(45.9,Spider-Realistic) (37.5,Spider-SYN) (39.6,Spider-DK)};

        \addplot [
            line width= .3mm, 
            fill=purple,
            postaction={
                pattern=north east lines, 
                pattern color=white
            }
        ]
        coordinates {(71.1,Spider-Realistic) (63.3,Spider-SYN) (61.7,Spider-DK)};

        \legend{\textsc{ChatGPT-SQL}, \textsc{C3}, \textsc{PURPLE}}
        \end{axis}
    \end{scope}

    \begin{scope}[xshift=3.8cm]
        \begin{axis}[
        xbar = .17mm, 
        height=4cm, 
        width=5cm,
            xmin = 0,
            xmax = 100,
            xtick={20,40,60,80},
            enlarge y limits=0.3, 
            axis line style=thick,
            xlabel={EX Score \%},
            xlabel style={font=\scriptsize, yshift=1mm}, 
            xticklabel style={font=\scriptsize},
            symbolic y coords={Spider-Realistic, Spider-SYN, Spider-DK}, 
            yticklabel style={font=\scriptsize, text opacity=0},
            ytick={Spider-Realistic, Spider-SYN, Spider-DK},
            bar width=5pt,
            xmajorgrids=true,
            ymajorgrids=true,
            grid style=dashed,
            nodes near coords align={vertical},
            legend image code/.code={%
                \draw[#1, draw=black] (0cm,-0.1cm) rectangle (0.28cm,0.05cm);
            }, 
            legend style={font=\tiny, at={(0.85,0.95)},
                anchor=north,
                row sep=-0.08cm,
            }
        ]

        \addplot [
            line width= .3mm, 
            fill=pink,
            postaction={
                pattern=north east lines, 
                pattern color=white
            }
        ]
        coordinates {(63.4,Spider-Realistic) (58.6,Spider-SYN) (62.6,Spider-DK)};
        
        \addplot [
            line width= .3mm, 
            fill=palegreen,
            postaction={
                pattern=north east lines, 
                pattern color=white
            }
        ]
        coordinates {(77.8,Spider-Realistic) (70.0,Spider-SYN) (67.5,Spider-DK)};

        \addplot [
            line width= .3mm, 
            fill=purple,
            postaction={
                pattern=north east lines, 
                pattern color=white
            }
        ]
        coordinates {(79.9,Spider-Realistic) (74.0,Spider-SYN) (75.3,Spider-DK)};
        \end{axis}
    \end{scope}  
    \end{tikzpicture}
    \caption{Comparison of EM/EX scores on Spider-DK, Spider-SYN and Spider-Realistic.}
    \label{fig:generalization_ability}
\end{figure}

As shown in Figure~\ref{fig:generalization_ability}, PURPLE achieves the best EM score, a notable achievement for LLM-based approaches that often struggle in this area. Specifically, PURPLE registers EM scores of 61.7\%, 63.3\%, and 71.1\% on Spider-DK, Spider-SYN, and Spider-Realistic benchmarks, respectively. These results are over 22\% better than C3, demonstrating a superior ability to generate accurate SQL compared to other methods.

Figure~\ref{fig:generalization_ability} also shows that PURPLE consistently maintains high EX scores across the three benchmarks, with 75.3\%, 74.0\%, and 79.9\% on Spider-DK, Spider-SYN, and Spider-Realistic, respectively. This uniformity in performance illustrates the robustness of PURPLE relative to its counterparts.

Although PURPLE incorporates fine-tuning for demonstration retrieval, it avoids a performance drop across varying data distributions. Because the fine-tuned model is utilized to enhance the operator composition knowledge as the intermediary.

\subsection{Cost \textit{v.s.} Performance}

PURPLE forms the prompt based on the token number to control the budget for each SQL translation. We evaluate the performance of PURPLE on the Spider under varying budget constraints, focusing on input length and the number of responses. We evaluate with input token limitation ($len$) of ${512, 1024, 2048, 3072}$ and consistency numbers ($num$) from ${1, 10, 20, 30, 40}$. The $num$ is to control the generated token number. Figure~\ref{fig:performance_budget_tradeoff} shows the accuracy under different budgets.

\begin{figure}
    \vspace{-2mm}
    \centering
    \definecolor{purple}{rgb}{0.36, 0.36, 0.5} 
    % \definecolor{purple}{rgb}{0.57, 0.55, 0.78}
    % \definecolor{gray}{rgb}{0.6, 0.6, 0.6}
    \definecolor{gray}{rgb}{0.7, 0.7, 0.7}
    \definecolor{darkgray}{rgb}{0.3, 0.3, 0.3}
    % \hspace*{-0.25cm}
    \begin{tikzpicture}[scale=0.5]
    \begin{scope}[xshift=-5.5cm]
    \draw[-latex]{(0.45, -4.5)--(0.45, -0.1)};
    \draw[-latex]{(0.45, -4.5)--(5.9, -4.5)};

    \node[minimum size=8mm, text=black, font=\tiny] at (0.1,0) {\textit{len}};
    \node[minimum size=8mm, text=black, font=\tiny] at (5.6,-4.8) {\textit{num}};
    
      \foreach \y [count=\n] in {
          {73.9,74.1,75.0,76.1,0.0},
          {74.0,74.0,74.6,75.7,74.5},
          {73.1,73.3,73.8,73.9,73.2},
          {63.5,63.9,63.7,64.8,64.5},
        } {
          \foreach \x [count=\m] in \y {
            \pgfmathparse{int(\x)}
            \let\r\pgfmathresult
            \ifnum\r>0{
                \pgfmathparse{(\x-60)*(\x-60)*(\x-60)/32+25}
                \let\c\pgfmathresult
                \node[fill=purple!\c!gray, minimum size=5mm, text=white, font=\tiny] at (\m,-\n) {\x};
            }
            \else{
                \node[fill=gray, minimum size=5mm, text=white, font=\tiny] at (\m,-\n) {N/A};
            }
            \fi
            
          }
        }
        \foreach \a [count=\i] in {3072,2048,1024,512} {
            \node[minimum size=5mm, font=\tiny] at (0, -\i) {\a};
        }
        \foreach \a [count=\i] in {1,10,20,30,40} {
            \node[minimum size=5mm, font=\tiny] at (\i, -4.9) {\a};
        }
        \node[font=\scriptsize] at (3,0) {EM Score \%};
    \end{scope}

    \begin{scope}
        
    \draw[-latex]{(0.45, -4.5)--(0.45, -0.1)};
    \draw[-latex]{(0.45, -4.5)--(5.9, -4.5)};
    
    \node[minimum size=8mm, text=black, font=\tiny] at (0.1,0) {\textit{len}};
    \node[minimum size=8mm, text=black, font=\tiny] at (5.6,-4.8) {\textit{num}};
    
      \foreach \y [count=\n] in {
            {83.4, 83.6, 84.6, 84.8, 0},
            {83.6, 84.4, 84.9, 84.6, 84.7},
            {83.5, 84.4, 84.0, 83.7, 84.6},
            {81.8, 82.2, 83.5, 83.2, 83.3},
        } {
          % heatmap tiles
          \foreach \x [count=\m] in \y {
            \pgfmathparse{int(\x)}
            \let\r\pgfmathresult
            \ifnum\r>0{
                \pgfmathparse{(\x-81)*38}
                \let\c\pgfmathresult
                \node[fill=purple!\c!gray, minimum size=5mm, text=white, font=\tiny] at (\m,-\n) {\x};
            }
            \else{
                \node[fill=gray, minimum size=5mm, text=white, font=\tiny] at (\m,-\n) {N/A};
            }
            \fi
            
          }
        }
    
        \foreach \a [count=\i] in {1,10,20,30,40} {
            \node[minimum size=6mm, font=\tiny] at (\i, -4.9) {\a};
        }
      \node[font=\scriptsize] at (3,0) {EX Score \%};
    \end{scope}

    \begin{scope}[xshift=5.5cm]
    \draw[-latex]{(0.45, -4.5)--(0.45, -0.1)};
    \draw[-latex]{(0.45, -4.5)--(5.9, -4.5)};
    
    \node[minimum size=8mm, text=black, font=\tiny] at (0.1,0) {\textit{len}};
    \node[minimum size=8mm, text=black, font=\tiny] at (5.6,-4.8) {\textit{num}};
    
      \foreach \y [count=\n] in {
            {3.04,3.30,3.58,3.86,0},
            {2.02,2.27,2.56,2.84,3.13},
            {1.00,1.25,1.54,1.83,2.11},
            {0.48,0.73,1.00,1.28,1.56},
        } {
        
          \foreach \x [count=\m] in \y {
            \pgfmathparse{int(\x+0.9)}
            \let\r\pgfmathresult
            \ifnum\r>0{
                \pgfmathparse{(\x*1000/30+25}
                \let\c\pgfmathresult
                \node[fill=purple!\c!gray, minimum size=5mm, text=white, font=\tiny] at (\m,-\n) {\x};
            }
            \else{
                \node[fill=gray, minimum size=5mm, text=white, font=\tiny] at (\m,-\n) {N/A};
            }
            \fi
            
          }
        }
    
        \foreach \a [count=\i] in {1,10,20,30,40} {
            \node[minimum size=6mm, font=\tiny] at (\i, -4.9) {\a};
        }
      \node[font=\scriptsize] at (3,0) {Token Cost/Query(K)};
    \end{scope}
    \end{tikzpicture}
    \vspace{-6mm}
    \caption{PURPLE (ChatGPT) performance and token consumption under various budget settings. $len$ represents prompt length, $num$ represents consistency number.}
    \label{fig:performance_budget_tradeoff}
    \vspace{-3mm}
\end{figure}

The performance of PURPLE tends to enhance with an increased budget. However, the increase becomes marginal when the input length surpasses 2,048 tokens. 
This is because adding more tokens offers diminishing returns on its ability to generalize.
Due to LLM limitations, a single call can process only up to 4,096 tokens, shown as N/A in Figure~\ref{fig:performance_budget_tradeoff}.

Our default configuration for PURPLE is set with an input length of 3,072 and a consistency number of 30. For context, DIN-SQL with GPT4 consumes roughly 10,000 tokens for each query translation. C3 uses about 8,000 tokens, splitting between 1,000 for input and 7,000 for output.
DAIL-SQL, which can adjust input and output length, works best with around 3,000 tokens. Meanwhile, PURPLE outperforms these with only 1,250 tokens in ChatGPT, highlighting its efficiency.

\subsection{Robustness of Demonstration Selection}\label{robustness}

We evaluate the robustness of our demonstration selection algorithm by varying the initial parameter $p_0$ and adjusting the {\fontfamily{pcr}\selectfont \textsc{Increase}-Generalization} method, as shown in Algorithm~\ref{algoritm:demonstration_selection}. We also explore how inaccuracies in skeleton prediction impact performance.

\begin{figure}[h]
\centering
\hspace*{-10pt}
\begin{tikzpicture}
\begin{groupplot}[
    group style={
        group size=2 by 2, % 2x2 subplot layout
        vertical sep=6pt, % Reduced vertical separation
        horizontal sep=6pt, % Reduced horizontal separation
    },
    width=0.6\columnwidth, % Adjusted width to better fit the column
    height=0.4\columnwidth, % Adjusted height for better aspect ratio
    grid=major,
    grid style={dashed,gray!30},
    title style={font=\scriptsize},
    label style={font=\scriptsize},
    tick label style={font=\tiny}, % Further reduced tick label size
    every axis plot/.append style={thin}, % Make plot lines thinner to save space
    title style={yshift=-1.5ex}, % Shift title up to save space
    ylabel style={yshift=-1ex}, %
    xlabel style={yshift=1ex}, % Adjust label position to save space
]

\nextgroupplot[
    ylabel={EM Score \%}, 
    xticklabels={}, 
    ymin=65, ymax=95, 
    ytick={70, 75, ..., 90},
    legend style={
        font=\tiny, 
        legend columns=-1,
        inner xsep=1pt, 
        inner ysep=0pt, 
        nodes={scale=0.75, transform shape}, 
    },
    legend pos=north east, 
]
\addplot[mark=triangle, mark options={darkred}, darkred, line width=0.8pt]  coordinates {(1, 76.1) (2, 74.7) (3, 73.0) (4, 73.4)};
\addlegendentry{Linear-1};

\addplot[mark=square, mark options={darkblue}, darkblue, line width=0.8pt]  coordinates {(1, 74.85) (2, 74.27) (3, 74.27) (4, 73.5)};
\addlegendentry{Linear-3};

\addplot[mark=pentagon, mark options={darkgreen}, darkgreen, line width=0.8pt] coordinates {(1, 75.24) (2, 75.44) (3, 74.47) (4, 74.08)};
\addlegendentry{Exp-2};

\nextgroupplot[
    xtick={1, 2, 3, 4}, 
    xticklabels={}, 
    ymin=65, ymax=95, 
    ytick={70, 75, ..., 90},
    yticklabels={}, 
    legend style={
        font=\tiny, 
        legend columns=-1,
        inner xsep=1pt, 
        inner ysep=0pt, 
        nodes={scale=0.75, transform shape}, 
    },
    legend pos=north east, 
]
\addplot[mark=triangle, mark options={darkred}, darkred, line width=0.8pt]  coordinates {(1, 76.1) (2, 73.98) (3, 73.21) (4, 68.76)};
\addlegendentry{Drop-0};

\addplot[mark=|, mark options={darkblue}, darkblue, line width=0.8pt]  coordinates {(1, 74.08) (2, 73.31) (3, 70.99) (4, 68.47)};
\addlegendentry{Drop-0.5};

\addplot[mark=x, mark options={darkgreen}, darkgreen, line width=0.8pt] coordinates {(1, 73.21) (2, 71.18) (3, 70.31) (4, 67.89)};
\addlegendentry{Drop-1};

\nextgroupplot[
    xlabel={$p_0$}, 
    ylabel={EX Score \%}, 
    xtick={1, 2, 3, 4}, 
    xticklabels={1, 3, 6, 9}, 
    ymin=65, ymax=95, 
    ytick={70, 75, ..., 90},
    legend style={
        font=\tiny, 
        legend columns=-1,
        inner xsep=1pt, 
        inner ysep=0pt, 
        nodes={scale=0.75, transform shape}, 
    },
    legend pos=north east, 
]
\addplot[mark=triangle, mark options={darkred}, darkred, line width=0.8pt] coordinates {(1, 84.8) (2, 83.85) (3, 84.43) (4, 84.24)};
\addlegendentry{Linear-1};

\addplot[mark=square, mark options={darkblue}, darkblue, line width=0.8pt] coordinates {(1, 85.2) (2, 84.53) (3, 84.7) (4, 84.4)};
\addlegendentry{Linear-3};

\addplot[mark=pentagon, mark options={darkgreen}, darkgreen, line width=0.8pt] coordinates {(1, 85.2) (2, 84.72) (3, 84.33) (4, 84.43)};
\addlegendentry{Exp-2};

\nextgroupplot[
    xlabel={\textit{Masking number}}, 
    xtick={1, 2, 3, 4}, 
    xticklabels={0, 1, 2, 3}, 
    ymin=65, ymax=95, 
    ytick={70, 75, ..., 90},
    yticklabels={},
    legend style={
        font=\tiny, 
        legend columns=-1,
        inner xsep=1pt, 
        inner ysep=0pt, 
        nodes={scale=0.75, transform shape}, 
    },
    legend pos=north east, 
]
\addplot[mark=triangle, mark options={darkred}, darkred, line width=0.8pt] coordinates {(1, 84.8) (2, 83.9) (3, 84.4) (4, 84.2)};
\addlegendentry{Drop-0};

\addplot[mark=|, mark options={darkblue}, darkblue, line width=0.8pt]coordinates {(1, 84.8) (2, 84.72) (3, 83.66) (4, 84.04)};
\addlegendentry{Drop-0.5};

\addplot[mark=x, mark options={darkgreen}, darkgreen, line width=0.8pt] coordinates {(1, 83.85) (2, 83.56) (3, 82.98) (4, 82.88)};
\addlegendentry{Drop-1};

\end{groupplot}
\end{tikzpicture}
\vspace{-3mm}
\caption{Robustness evaluation for demonstration selection}
\label{fig:robustness_evaluation}
\end{figure}
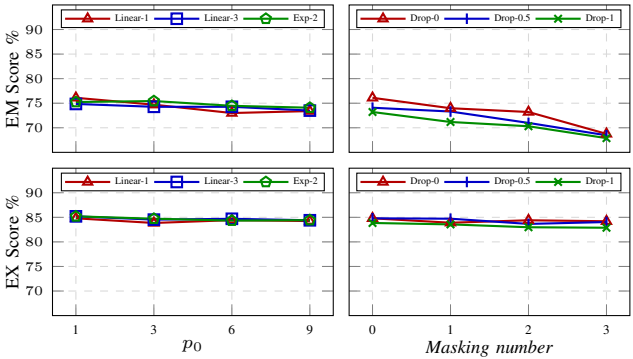

The left side of Figure~\ref{fig:robustness_evaluation} shows the performance of PURPLE with various $p_0$ and generalization methods. For example, \textit{Linear-1(3)} adds 1(3) to $p$ at each step, and \textit{Exp-2} doubles $p$ at each step. We found that the changes in performance are minor, less than 3\% for EM and less than 1.5\% for EX. This indicates that the performance of PURPLE is relatively stable.

On the right side of Figure~\ref{fig:robustness_evaluation}, we evaluate the effects of inaccurate skeletons on PURPLE. To simulate inaccuracies, we introduced two types of noise: $masking \ number=x$ simulates missing detailed information by ignoring the first $x$ layers of automaton abstraction, and $Drop-y$ randomly drops one predicted skeleton with a probability of $y$. For instance, $Drop-0.5$ and $masking \ number=2$ drops one predicted skeleton half the time and ignores the first two abstraction levels (Detail-Level and Keywords-Level) during demonstration selection. We observed a drop in EM scores with more noise. However, even in tough scenarios like only using Clause-Level information, PURPLE still achieves competitive EM scores, demonstrating its resilience to prediction inaccuracies.

\subsection{Performance with Various LLMs}\label{various_llms}

For the LLMs-based approaches, the selection of the specific LLM can lead to variations in performance. We evaluate the performance variations on the Spider of DIN-SQL, C3, DAIL-SQL, and PURPLE when utilizing either ChatGPT or GPT4. The results are shown in Table~\ref{tab:llm_robustness}.

\begin{table}[h]\footnotesize
    \centering
    \caption{EM/EX comparison between ChatGPT and GPT4.}
    \begin{tabular}{p{1.5cm}  p{1.2cm}  p{1.3cm}<{\centering}  p{1.3cm}<{\centering}}
        \hline
        \textbf{Strategy}       & \textbf{LLM}  & \textbf{EM\%} & \textbf{EX\%} \\ 
        \hline
        \hline
        \multirow{2}*{DIN-SQL}  & GPT4          &     60.1        &     82.8       \\
                                & ChatGPT       &     43.0(\textcolor{darkred}{-17.1}) &     75.5(\textcolor{darkred}{-7.3}) \\
        
        \hline
        \multirow{2}*{C3}      & GPT4           &      50.7       &     82.1       \\
                               & ChatGPT        &      43.1(-7.6) &     81.8(-0.3) \\
        \hline
        \multirow{2}*{DAIL-SQL} & GPT4          &      68.7       &     83.6       \\
                                & ChatGPT       &      65.1(-3.6) &     81.3(-2.3) \\
        \hline
        \multirow{2}*{PURPLE}  & GPT4           &      80.5       &     87.8      \\
                               & ChatGPT        &      76.1(-4.4) &     84.8(-3.0) \\
        \hline
        \end{tabular}
    
    \label{tab:llm_robustness}
    
\end{table}

PURPLE consistently outperforms others, whether using ChatGPT or GPT4. Meanwhile, DIN-SQL exhibits a sensitivity to the LLMs. DIN-SQL employs CoT methodology, which relies on the reasoning capabilities inherent in GPT4. ChatGPT struggles with complex reasoning tasks, thereby increasing the risk of error propagation~\cite{error_propagation}. 
The sensitivity to the LLMs not only raises its cost but also undermines its robustness.

C3 shows stable performance across both LLMs. However, it underutilizes the capabilities of GPT4. The hand-crafted instructions limit its SQL knowledge. Lacking operator composition knowledge in prompt restricts the potential enhancements.

DAIL-SQL exhibits a parallel trend in performance variability between GPT4 and ChatGPT, similar to PURPLE. Because they both propose to utilize an adaptable demonstration selection strategy. However, DAIL-SQL lacks sufficient operator composition knowledge to achieve better performance.

PURPLE consistently outperforms other strategies with both ChatGPT and GPT4. It stays accurate in tight resource settings, showing the importance of giving LLMs the logical operator composition knowledge they need for tasks.

\subsection{Ablation Study}

We conduct an ablation study to show the contribution of each module in PURPLE. 
The results are shown in Table~\ref{tab:ablation_study}.

\begin{table}[h]\footnotesize
    \centering
    \caption{Ablation Study.}
    \begin{tabular}{p{3.0cm} p{1.3cm}<{\centering} p{1.3cm}<{\centering}}
        \hline
        \textbf{Strategy}        & \textbf{EM\%} & \textbf{EX\%} \\ 
        % \hline
        \hline
        PURPLE (ChatGPT)          &     76.1 &     84.8 \\
        % \hline
        \ -Schema Pruning        &     71.2(-4.9)       &     83.4(-1.4) \\
        \ -Steiner Tree        &     75.0(-1.1)       &     84.4(-0.3) \\
        \ -Demonstration Selection   &     59.1(-17.0)  &     81.6(-3.2) \\
        \ -Database Adaption     &     74.7(-1.4)       &     81.8(-3.0) \\
        \hline
        +Oracle Skeleton         &     78.8(+2.7)       &     86.8(+2.0) \\
        \hline
        \end{tabular}
    \label{tab:ablation_study}
    \vspace{-3mm}
\end{table}

The schema pruning module simplifies tasks, helping LLMs focus on key information for SQL generation. The EM and EX scores suffer from a large drop without such a module (-Schema Pruning). 
To evaluate the effectiveness of the Steiner Tree-based pruning strategy, we compared it with the pruning approach used by RESDSQL (-Steiner Tree). Both the EM and EX scores are lower when using the RESDSQL method. This is because it requires LLMs to process more information, highlighting the efficiency of our approach.

Demonstration selection is the key module of PURPLE. Randomly selecting demonstrations (-Demonstration Selection) greatly reduced EM scores, showing the importance of composition knowledge for the SQL generation of LLMs.

The database adaptation module further builds upon the successes of the previous modules, contributing to the stabilization of model outputs and mitigating hallucination issues.

Additionally, we conducted an oracle-setting experiment. We replace the predicted skeletons with the oracle skeleton. PURPLE with ChatGPT (+Oracle Skeleton) achieves an EM score of 78.8\% and an EX score of 86.8\%. 
This result highlights the importance of accurately predicting logical composition knowledge in the overall performance of PURPLE. Improving skeleton prediction could further boost results.

\section{Related Works}\label{sec:related_works}

The NL2SQL task has been under investigation for decades with the advancements of NLP. Early studies like~\cite{DBLP:E1, DBLP:E2, DBLP:E3, DBLP:E5, DBLP:E4, DBLP:E6, DBLP:E7, DBLP:E8} on NL2SQL mainly focus on rule-based mapping, which has limited generalization ability. Modern NL2SQL approaches incorporate SOTA models to optimize performance. We classify the LMs-based NL2SQL approaches into PLMs-based and LLMs-based NL2SQL as shown in Section~\ref{sec:preliminaries}.

\textbf{PLMs-based NL2SQL.} Fine-tuning PLMs as part of a sequence-to-sequence paradigm is one of the popular approaches for NL2SQL. Such fine-tuning techniques empower the development of custom modules on the foundation of PLMs. Works like RAT-SQL~\cite{DBLP:RAT-SQL}, GNN~\cite{DBLP:gnn_text2sql}, GlobalGNN~\cite{DBLP:global_gnn_text2sql} BIRDGE~\cite{DBLP:BRIDGE}, LGESQL~\cite{DBLP:lgesql}, $S^2$SQL~\cite{DBLP:S2SQL}, RASAT~\cite{DBLP:RASAT} and Graphix-T5~\cite{DBLP:GRAPHIX} have introduced novel encoder architectures for enhanced semantic comprehension. IRNet~\cite{DBLP:intermediate_representation}, SmBoP~\cite{DBLP:smbop}, NatSQL~\cite{DBLP:NATSQL}, PICARD~\cite{DBLP:PICARD}, CATSQL~\cite{DBLP:CATSQL}, and SC-Prompt~\cite{DBLP:SCPrompt} focus on reducing decoder complexity, while others such as RESDSQL~\cite{DBLP:RESD}, N-best~\cite{DBLP:N-best}, GAR~\cite{DBLP:GAR}, GenSQL~\cite{DBLP:GenSQL} and MetaSQL~\cite{MetaSQL} integrate ranking models to elevate performance. However, the PLMs-based approaches are constrained by model and pre-trained corpus sizes, leading to misunderstanding of user intentions. As models grow, their adaptability for NL2SQL tasks diminishes, prompting a shift of research attention towards LLMs-based methodologies.

\textbf{LLMs-based NL2SQL.} Techniques for integrating LLMs into NL2SQL can be categorized according to whether demonstrations are employed in the prompt, leading to categorizations as either zero-shot or few-shot methodologies. Zero-shot approaches, such as those explored in~\cite{DBLP:ChatGPT-zero-shot, DBLP:C3, DBLP:how_to_prompt, DBLP:ZERO}, aim to refine translation precision through instruction design.
Few-shot methodologies~\cite{DBLP:Synchromesh, DBLP:divide_and_prompt, DBLP:DIN-SQL, DBLP:ICL-SQL, DBLP:exploring_cot, DBLP:SKILL, DBLP:SQL-PALM, DBLP:revision_chain, DBLP:adapt_and_decompose}, contains related knowledge for teaching the LLMs about how to handle the translation task on hand.
For instance, DAIL-SQL~\cite{DBLP:DAIL-SQL}, SYNCHROMESH~\cite{DBLP:Synchromesh} and Linyong et al.\cite{DBLP:ICL-SQL} prioritize SQL-aligned demonstrations, while SKILL-KNN\cite{DBLP:SKILL} prefers to include semantically similar ones. Some propose retrieving demonstrations with the coverage for the prompt~\cite{DBLP:adapt_and_decompose}. CoT is an LLMs-based technique that has been applied on NL2SQL~\cite{DBLP:divide_and_prompt, DBLP:DIN-SQL, DBLP:exploring_cot}, which identifies that variations in CoT style can influence performance outcomes. Several multi-turn approaches~\cite{DBLP:SKILL, DBLP:revision_chain} propose to refine the generated SQL in multi-turn interactions with LLMs, achieving higher accuracy while suffering from high API cost. A common shortcoming among existing LLMs-based approaches is their inability to achieve high EM scores, often attributed to the challenge of controlling the generation process. 
In response to the observed limitations of LLMs in SQL writing,
PURPLE focuses on extracting essential logical operator composition knowledge for logical enhancement, leading to a new SOTA performance.

\section{Conclusion and Future Work}\label{sec:conclusion_and_future_work}

We introduced PURPLE, a novel LLMs-based NL2SQL approach that enhanced translation precision through demonstration selection. 
PURPLE models the operator composition knowledge by a four-level automaton, and related automaton construction and matching strategies are designed for demonstration selection.
Schema pruning and skeleton prediction facilitate this selection process,
and the database adaptation module serves to stabilize outputs and mitigate hallucination issues. 
PURPLE successfully enhanced the LLMs with SQL operator composition knowledge, achieving reliable performance on four popular benchmarks. 
We also evaluated the robustness and the influence of LLM selection for PURPLE.

One promising research direction is the development of generation-based prompting methods. While PURPLE effectively retrieves existing demonstrations to construct prompts, this retrieval-based strategy is inherently limited by the available pool of demonstrations. Involving generating prompts directly using PLMs is a potentially more flexible approach. This method could offer a more generalized and intuitive way to create prompts. However, the primary challenge with a generation-based approach lies in fine-tuning PLMs to produce optimized prompts effectively. Although there has been some success with using reinforcement learning for prompt optimization in prior research~\cite{DBLP:RLPrompt, DBLP:DynamicPrompt}, fine-tuning PLMs specifically for prompt generation presents difficulties. Using existing demonstrations as a basis, like PURPLE, could serve as a valuable starting point for developing more advanced generation-based prompting methods in the future.

\section{Acknowledgments}

The authors thank the reviewers for the valuable feedback. This work was supported by NSFC (Grant No. 62272106). The corresponding authors are Zhenying He and X.Sean Wang.

\bibliographystyle{IEEEtran}
\bibliography{ref}

\begin{thebibliography}{10}
\providecommand{\url}[1]{#1}
\csname url@samestyle\endcsname
\providecommand{\newblock}{\relax}
\providecommand{\bibinfo}[2]{#2}
\providecommand{\BIBentrySTDinterwordspacing}{\spaceskip=0pt\relax}
\providecommand{\BIBentryALTinterwordstretchfactor}{4}
\providecommand{\BIBentryALTinterwordspacing}{\spaceskip=\fontdimen2\font plus
\BIBentryALTinterwordstretchfactor\fontdimen3\font minus \fontdimen4\font\relax}
\providecommand{\BIBforeignlanguage}[2]{{%
\expandafter\ifx\csname l@#1\endcsname\relax
\typeout{** WARNING: IEEEtran.bst: No hyphenation pattern has been}%
\typeout{** loaded for the language `#1'. Using the pattern for}%
\typeout{** the default language instead.}%
\else
\language=\csname l@#1\endcsname
\fi
#2}}
\providecommand{\BIBdecl}{\relax}
\BIBdecl

\bibitem{DBLP:Synchromesh}
G.~Poesia, A.~Polozov, V.~Le, A.~Tiwari, G.~Soares, C.~Meek, and S.~Gulwani, ``Synchromesh: Reliable code generation from pre-trained language models,'' in \emph{ICLR}, 2022.

\bibitem{DBLP:DIN-SQL}
M.~Pourreza and D.~Rafiei, ``{DIN-SQL:} decomposed in-context learning of text-to-sql with self-correction,'' \emph{CoRR}, 2023.

\bibitem{DBLP:SQL-PALM}
R.~Sun, S.~{\"{O}}. Arik, H.~Nakhost, H.~Dai, R.~Sinha, P.~Yin, and T.~Pfister, ``Sql-palm: Improved large language model adaptation for text-to-sql,'' \emph{CoRR}, 2023.

\bibitem{DBLP:EvaluatingLLM}
N.~Rajkumar, R.~Li, and D.~Bahdanau, ``Evaluating the text-to-sql capabilities of large language models,'' \emph{CoRR}, 2022.

\bibitem{DBLP:ChatGPT-zero-shot}
A.~Liu, X.~Hu, L.~Wen, and P.~S. Yu, ``A comprehensive evaluation of chatgpt's zero-shot text-to-sql capability,'' \emph{CoRR}, 2023.

\bibitem{DBLP:SELF-DEBUG}
X.~Chen, M.~Lin, N.~Sch{\"{a}}rli, and D.~Zhou, ``Teaching large language models to self-debug,'' \emph{CoRR}, 2023.

\bibitem{DBLP:ZERO}
Z.~Gu, J.~Fan, N.~Tang, S.~Zhang, Y.~Zhang, Z.~Chen, L.~Cao, G.~Li, S.~Madden, and X.~Du, ``Interleaving pre-trained language models and large language models for zero-shot {NL2SQL} generation,'' \emph{CoRR}, 2023.

\bibitem{DBLP:DAIL-SQL}
D.~Gao, H.~Wang, Y.~Li, X.~Sun, Y.~Qian, B.~Ding, and J.~Zhou, ``Text-to-sql empowered by large language models: {A} benchmark evaluation,'' \emph{CoRR}, 2023.

\bibitem{DBLP:COT}
J.~Wei, X.~Wang, D.~Schuurmans, M.~Bosma, B.~Ichter, F.~Xia, E.~H. Chi, Q.~V. Le, and D.~Zhou, ``Chain-of-thought prompting elicits reasoning in large language models,'' in \emph{NeurIPS}, 2022.

\bibitem{DBLP:Spider}
T.~Yu, R.~Zhang, K.~Yang, M.~Yasunaga, D.~Wang, Z.~Li, J.~Ma, I.~Li, Q.~Yao, S.~Roman, Z.~Zhang, and D.~R. Radev, ``Spider: {A} large-scale human-labeled dataset for complex and cross-domain semantic parsing and text-to-sql task,'' in \emph{EMNLP}, 2018,  3911--3921.

\bibitem{DBLP:C3}
X.~Dong, C.~Zhang, Y.~Ge, Y.~Mao, Y.~Gao, L.~Chen, J.~Lin, and D.~Lou, ``{C3:} zero-shot text-to-sql with chatgpt,'' \emph{CoRR}, 2023.

\bibitem{DBLP:empowerd}
D.~Gao, H.~Wang, Y.~Li, X.~Sun, Y.~Qian, B.~Ding, and J.~Zhou, ``Text-to-sql empowered by large language models: {A} benchmark evaluation,'' \emph{CoRR}, 2023.

\bibitem{DBLP:llm_few_shot_learner}
T.~B. Brown, B.~Mann, N.~Ryder, M.~Subbiah, J.~Kaplan, P.~Dhariwal, A.~Neelakantan, P.~Shyam, G.~Sastry, A.~Askell, S.~Agarwal, A.~Herbert{-}Voss, G.~Krueger, T.~Henighan, R.~Child, A.~Ramesh, D.~M. Ziegler, J.~Wu, C.~Winter, C.~Hesse, M.~Chen, E.~Sigler, M.~Litwin, S.~Gray, B.~Chess, J.~Clark, C.~Berner, S.~McCandlish, A.~Radford, I.~Sutskever, and D.~Amodei, ``Language models are few-shot learners,'' in \emph{NeurIPS}, 2020.

\bibitem{DBLP:BERT}
J.~Devlin, M.~Chang, K.~Lee, and K.~Toutanova, ``{BERT:} pre-training of deep bidirectional transformers for language understanding,'' in \emph{NAACL}, 2019,  4171--4186.

\bibitem{DBLP:BART}
M.~Lewis, Y.~Liu, N.~Goyal, M.~Ghazvininejad, A.~Mohamed, O.~Levy, V.~Stoyanov, and L.~Zettlemoyer, ``{BART:} denoising sequence-to-sequence pre-training for natural language generation, translation, and comprehension,'' \emph{CoRR}, 2019.

\bibitem{DBLP:T5}
C.~Raffel, N.~Shazeer, A.~Roberts, K.~Lee, S.~Narang, M.~Matena, Y.~Zhou, W.~Li, and P.~J. Liu, ``Exploring the limits of transfer learning with a unified text-to-text transformer,'' \emph{J. Mach. Learn. Res.}, vol.~21,  140:1--140:67, 2020.

\bibitem{DBLP:GPT3}
T.~B. Brown, B.~Mann, N.~Ryder, M.~Subbiah, J.~Kaplan, P.~Dhariwal, A.~Neelakantan, P.~Shyam, G.~Sastry, A.~Askell, S.~Agarwal, A.~Herbert{-}Voss, G.~Krueger, T.~Henighan, R.~Child, A.~Ramesh, D.~M. Ziegler, J.~Wu, C.~Winter, C.~Hesse, M.~Chen, E.~Sigler, M.~Litwin, S.~Gray, B.~Chess, J.~Clark, C.~Berner, S.~McCandlish, A.~Radford, I.~Sutskever, and D.~Amodei, ``Language models are few-shot learners,'' \emph{CoRR}, 2020.

\bibitem{DBLP:PALM}
A.~Chowdhery, S.~Narang, J.~Devlin, M.~Bosma, G.~Mishra, A.~Roberts, P.~Barham, H.~W. Chung, C.~Sutton, S.~Gehrmann, P.~Schuh, K.~Shi, S.~Tsvyashchenko, J.~Maynez, A.~Rao, P.~Barnes, Y.~Tay, N.~Shazeer, V.~Prabhakaran, E.~Reif, N.~Du, B.~Hutchinson, R.~Pope, J.~Bradbury, J.~Austin, M.~Isard, G.~Gur{-}Ari, P.~Yin, T.~Duke, A.~Levskaya, S.~Ghemawat, S.~Dev, H.~Michalewski, X.~Garcia, V.~Misra, K.~Robinson, L.~Fedus, D.~Zhou, D.~Ippolito, D.~Luan, H.~Lim, B.~Zoph, A.~Spiridonov, R.~Sepassi, D.~Dohan, S.~Agrawal, M.~Omernick, A.~M. Dai, T.~S. Pillai, M.~Pellat, A.~Lewkowycz, E.~Moreira, R.~Child, O.~Polozov, K.~Lee, Z.~Zhou, X.~Wang, B.~Saeta, M.~Diaz, O.~Firat, M.~Catasta, J.~Wei, K.~Meier{-}Hellstern, D.~Eck, J.~Dean, S.~Petrov, and N.~Fiedel, ``Palm: Scaling language modeling with pathways,'' \emph{CoRR}, 2022.

\bibitem{DBLP:BRIDGE}
X.~V. Lin, R.~Socher, and C.~Xiong, ``Bridging textual and tabular data for cross-domain text-to-sql semantic parsing,'' in \emph{EMNLP}, 2020,  4870--4888.

\bibitem{DBLP:consistency}
C.~Zhou, J.~He, X.~Ma, T.~Berg{-}Kirkpatrick, and G.~Neubig, ``Prompt consistency for zero-shot task generalization,'' in \emph{EMNLP}, 2022,  2613--2626.

\bibitem{DBLP:PICARD}
T.~Scholak, N.~Schucher, and D.~Bahdanau, ``{PICARD:} parsing incrementally for constrained auto-regressive decoding from language models,'' in \emph{EMNLP}, 2021,  9895--9901.

\bibitem{DBLP:RESD}
H.~Li, J.~Zhang, C.~Li, and H.~Chen, ``{RESDSQL:} decoupling schema linking and skeleton parsing for text-to-sql,'' in \emph{AAAI}, 2023,  13\,067--13\,075.

\bibitem{DBLP:focal_loss}
T.~Lin, P.~Goyal, R.~B. Girshick, K.~He, and P.~Doll{\'{a}}r, ``Focal loss for dense object detection,'' in \emph{ICCV}, 2017,  2999--3007.

\bibitem{DBLP:steiner_tree}
F.~K. Hwang and D.~S. Richards, ``Steiner tree problems,'' \emph{Networks}, vol.~22, no.~1,  55--89, 1992.

\bibitem{DBLP:DISCOVER}
V.~Hristidis and Y.~Papakonstantinou, ``{DISCOVER:} keyword search in relational databases,'' in \emph{VLDB}, 2002,  670--681.

\bibitem{DBLP:STP}
I.~Ljubic, ``Solving steiner trees: Recent advances, challenges, and perspectives,'' \emph{Networks}, vol.~77, no.~2,  177--204, 2021.

\bibitem{DBLP:beam_search}
M.~Freitag and Y.~Al{-}Onaizan, ``Beam search strategies for neural machine translation,'' in \emph{ACL}, T.~Luong, A.~Birch, G.~Neubig, and A.~M. Finch, Eds., 2017,  56--60.

\bibitem{DBLP:SPIDER-DK}
Y.~Gan, X.~Chen, and M.~Purver, ``Exploring underexplored limitations of cross-domain text-to-sql generalization,'' in \emph{EMNLP}, 2021,  8926--8931.

\bibitem{DBLP:SPIDER-SYN}
Y.~Gan, X.~Chen, Q.~Huang, M.~Purver, J.~R. Woodward, J.~Xie, and P.~Huang, ``Towards robustness of text-to-sql models against synonym substitution,'' in \emph{ACL}, 2021,  2505--2515.

\bibitem{DBLP:SPIDER-REALISTIC}
X.~Deng, A.~H. Awadallah, C.~Meek, O.~Polozov, H.~Sun, and M.~Richardson, ``Structure-grounded pretraining for text-to-sql,'' in \emph{NAACL}, 2021,  1337--1350.

\bibitem{DBLP:TEST-SUITE}
R.~Zhong, T.~Yu, and D.~Klein, ``Semantic evaluation for text-to-sql with distilled test suites,'' in \emph{EMNLP}, 2020,  396--411.

\bibitem{DBLP:RASAT}
J.~Qi, J.~Tang, Z.~He, X.~Wan, Y.~Cheng, C.~Zhou, X.~Wang, Q.~Zhang, and Z.~Lin, ``{RASAT:} integrating relational structures into pretrained seq2seq model for text-to-sql,'' in \emph{EMNLP}, 2022,  3215--3229.

\bibitem{DBLP:GRAPHIX}
J.~Li, B.~Hui, R.~Cheng, B.~Qin, C.~Ma, N.~Huo, F.~Huang, W.~Du, L.~Si, and Y.~Li, ``Graphix-t5: Mixing pre-trained transformers with graph-aware layers for text-to-sql parsing,'' in \emph{AAAI}, 2023,  13\,076--13\,084.

\bibitem{error_propagation}
S.~Yao, J.~Zhao, D.~Yu, N.~Du, I.~Shafran, K.~Narasimhan, and Y.~Cao, ``{ReAct}: Synergizing reasoning and acting in language models,'' in \emph{ICLR}, 2023.

\bibitem{DBLP:E1}
J.~M. Zelle and R.~J. Mooney, ``Learning to parse database queries using inductive logic programming,'' in \emph{AAAI}, 1996,  1050--1055.

\bibitem{DBLP:E2}
A.~Simitsis, G.~Koutrika, and Y.~E. Ioannidis, ``Pr{\'{e}}cis: from unstructured keywords as queries to structured databases as answers,'' \emph{VLDBJ}, vol.~17, no.~1,  117--149, 2008.

\bibitem{DBLP:E3}
F.~Li and H.~V. Jagadish, ``Constructing an interactive natural language interface for relational databases,'' \emph{VLDB}, vol.~8, no.~1,  73--84, 2014.

\bibitem{DBLP:E5}
D.~Saha, A.~Floratou, K.~Sankaranarayanan, U.~F. Minhas, A.~R. Mittal, and F.~{\"{O}}zcan, ``{ATHENA:} an ontology-driven system for natural language querying over relational data stores,'' \emph{VLDB}, vol.~9, no.~12,  1209--1220, 2016.

\bibitem{DBLP:E4}
F.~Li and H.~V. Jagadish, ``Nalir: an interactive natural language interface for querying relational databases,'' in \emph{SIGMOD}, 2014,  709--712.

\bibitem{DBLP:E6}
J.~Sen, C.~Lei, A.~Quamar, F.~{\"{O}}zcan, V.~Efthymiou, A.~Dalmia, G.~Stager, A.~R. Mittal, D.~Saha, and K.~Sankaranarayanan, ``{ATHENA++:} natural language querying for complex nested {SQL} queries,'' \emph{VLDB}, vol.~13, no.~11,  2747--2759, 2020.

\bibitem{DBLP:E7}
H.~Kim, B.~So, W.~Han, and H.~Lee, ``Natural language to {SQL:} where are we today?'' \emph{VLDB}, vol.~13, no.~10,  1737--1750, 2020.

\bibitem{DBLP:E8}
O.~Gkini, T.~Belmpas, G.~Koutrika, and Y.~E. Ioannidis, ``An in-depth benchmarking of text-to-sql systems,'' in \emph{SIGMOD}, 2021,  632--644.

\bibitem{DBLP:RAT-SQL}
B.~Wang, R.~Shin, X.~Liu, O.~Polozov, and M.~Richardson, ``{RAT-SQL:} relation-aware schema encoding and linking for text-to-sql parsers,'' in \emph{ACL}, 2020,  7567--7578.

\bibitem{DBLP:gnn_text2sql}
B.~Bogin, J.~Berant, and M.~Gardner, ``Representing schema structure with graph neural networks for text-to-sql parsing,'' in \emph{ACL}, 2019,  4560--4565.

\bibitem{DBLP:global_gnn_text2sql}
B.~Bogin, M.~Gardner, and J.~Berant, ``Global reasoning over database structures for text-to-sql parsing,'' in \emph{EMNLP}, 2019,  3657--3662.

\bibitem{DBLP:lgesql}
R.~Cao, L.~Chen, Z.~Chen, Y.~Zhao, S.~Zhu, and K.~Yu, ``{LGESQL:} line graph enhanced text-to-sql model with mixed local and non-local relations,'' in \emph{ACL}, 2021,  2541--2555.

\bibitem{DBLP:S2SQL}
B.~Hui, R.~Geng, L.~Wang, B.~Qin, Y.~Li, B.~Li, J.~Sun, and Y.~Li, ``S{\({^2}\)}sql: Injecting syntax to question-schema interaction graph encoder for text-to-sql parsers,'' in \emph{ACL}, 2022,  1254--1262.

\bibitem{DBLP:intermediate_representation}
J.~Guo, Z.~Zhan, Y.~Gao, Y.~Xiao, J.~Lou, T.~Liu, and D.~Zhang, ``Towards complex text-to-sql in cross-domain database with intermediate representation,'' in \emph{ACL}, 2019,  4524--4535.

\bibitem{DBLP:smbop}
O.~Rubin and J.~Berant, ``Smbop: Semi-autoregressive bottom-up semantic parsing,'' in \emph{ACL}, 2021,  12--21.

\bibitem{DBLP:NATSQL}
Y.~Gan, X.~Chen, J.~Xie, M.~Purver, J.~R. Woodward, J.~H. Drake, and Q.~Zhang, ``Natural {SQL:} making {SQL} easier to infer from natural language specifications,'' in \emph{EMNLP}, 2021,  2030--2042.

\bibitem{DBLP:CATSQL}
H.~Fu, C.~Liu, B.~Wu, F.~Li, J.~Tan, and J.~Sun, ``Catsql: Towards real world natural language to {SQL} applications,'' \emph{VLDB}, vol.~16, no.~6,  1534--1547, 2023.

\bibitem{DBLP:SCPrompt}
Z.~Gu, J.~Fan, N.~Tang, L.~Cao, B.~Jia, S.~Madden, and X.~Du, ``Few-shot text-to-sql translation using structure and content prompt learning,'' \emph{SIGMOD}, vol.~1, no.~2,  147:1--147:28, 2023.

\bibitem{DBLP:N-best}
L.~Zeng, S.~H.~K. Parthasarathi, and D.~Hakkani{-}Tur, ``N-best hypotheses reranking for text-to-sql systems,'' in \emph{SLT}, 2022,  663--670.

\bibitem{DBLP:GAR}
Y.~Fan, Z.~He, T.~Ren, D.~Guo, L.~Chen, R.~Zhu, G.~Chen, Y.~Jing, K.~Zhang, and X.~S. Wang, ``Gar: {A} generate-and-rank approach for natural language to {SQL} translation,'' in \emph{ICDE}, 2023,  110--122.

\bibitem{DBLP:GenSQL}
Y.~Fan, T.~Ren, Z.~He, X.~S. Wang, Y.~Zhang, and X.~Li, ``Gensql: {A} generative natural language interface to database systems,'' in \emph{ICDE}, 2023,  3603--3606.

\bibitem{MetaSQL}
Y.~Fan, Z.~He, T.~Ren, C.~Huang, Y.~Jing, K.~Zhang, and X.~S. Wang, ``Metasql: A generate-then-rank framework for natural language to sql translation,'' \emph{CoRR}, 2024.

\bibitem{DBLP:how_to_prompt}
S.~Chang and E.~Fosler{-}Lussier, ``How to prompt llms for text-to-sql: {A} study in zero-shot, single-domain, and cross-domain settings,'' \emph{CoRR}, 2023.

\bibitem{DBLP:divide_and_prompt}
X.~Liu and Z.~Tan, ``Divide and prompt: Chain of thought prompting for text-to-sql,'' \emph{CoRR}, 2023.

\bibitem{DBLP:ICL-SQL}
L.~Nan, Y.~Zhao, W.~Zou, N.~Ri, J.~Tae, E.~Zhang, A.~Cohan, and D.~Radev, ``Enhancing few-shot text-to-sql capabilities of large language models: {A} study on prompt design strategies,'' \emph{CoRR}, 2023.

\bibitem{DBLP:exploring_cot}
C.~Tai, Z.~Chen, T.~Zhang, X.~Deng, and H.~Sun, ``Exploring chain-of-thought style prompting for text-to-sql,'' \emph{CoRR}, 2023.

\bibitem{DBLP:SKILL}
S.~An, B.~Zhou, Z.~Lin, Q.~Fu, B.~Chen, N.~Zheng, W.~Chen, and J.~Lou, ``Skill-based few-shot selection for in-context learning,'' \emph{CoRR}, 2023.

\bibitem{DBLP:revision_chain}
C.~Guo, Z.~Tian, J.~Tang, S.~Li, Z.~Wen, K.~Wang, and T.~Wang, ``Retrieval-augmented gpt-3.5-based text-to-sql framework with sample-aware prompting and dynamic revision chain,'' \emph{CoRR}, 2023.

\bibitem{DBLP:adapt_and_decompose}
A.~Arora, S.~Bhaisaheb, H.~Nigam, M.~S. Patwardhan, L.~Vig, and G.~Shroff, ``Adapt and decompose: Efficient generalization of text-to-sql via domain adapted least-to-most prompting,'' \emph{CoRR}, 2023.

\bibitem{DBLP:RLPrompt}
M.~Deng, J.~Wang, C.~Hsieh, Y.~Wang, H.~Guo, T.~Shu, M.~Song, E.~P. Xing, and Z.~Hu, ``Rlprompt: Optimizing discrete text prompts with reinforcement learning,'' in \emph{EMNLP}, 2022,  3369--3391.

\bibitem{DBLP:DynamicPrompt}
P.~Lu, L.~Qiu, K.~Chang, Y.~N. Wu, S.~Zhu, T.~Rajpurohit, P.~Clark, and A.~Kalyan, ``Dynamic prompt learning via policy gradient for semi-structured mathematical reasoning,'' in \emph{ICLR}, 2023.

\end{thebibliography}

\end{document}